# Chemical templates that assemble the metal superhydrides

*Yuanhui Sun and Maosheng Miao**

Department of Chemistry and Biochemistry, California State University Northridge, CA USA

**Email: mmiao@csun.edu*

**The recent discoveries of many metal superhydrides provide a new route to room temperature superconductors. However, their stability and structure trends and the large chemical driving force needed to dissociate $H_2$ molecules and form H covalent network cannot be explained by direct metal-hydrogen bonds and volume effect. Here, we demonstrate that the understanding of superhydrides formation needs a perspective beyond traditional chemical bond theory. Using high-throughput calculations, we show that, after removing H atoms, the remaining metal lattices exhibit large electron localizations at the interstitial regions, which matches excellently to the H lattice like a template. Furthermore, H lattices consist of 3D aromatic building units that are greatly stabilized by chemical templates of metals close to s-d border. The chemical template theory can naturally explain the stability and structure trends of superhydrides and help predicting new materials such as two-metal superhydrides.**

Among many remarkable physical and chemical properties of hydrogen, its capability of achieving superconducting state at or above room temperature has received intensive attention in condensed matter studies for many decades.[1–3] However, hydrogen is very resistive to metallization and polymerization that is an essential step toward superconductivity, and large driving forces are needed to overcome it. As predicted theoretically, a tremendous high pressure of 550 GPa is needed to drive the transformation of hydrogen to atomic phases.[1,4,5] Therefore, the chemical forces that can "pre-compress" the hydrogen became an attractive approach of superconductive hydrogen under moderate pressure[2] and have led to the predictions and syntheses of numerous hydrogen-rich compounds that brought us ever close to room temperature superconductivity.[6,7] These compounds can be roughly categorized into metal hydrides and non-metal hydrides. As the second type of hydrides, $H_3S$ showed a $T_c$ of 203 K near 150 GPa,[8] which is the first to break the 25-year $T_c$ record of cuperates. More profoundly, a ternary C-S-H system has just been demonstrated to become superconducting at about 15 °C and 267 GPa.[9] All H atoms in these compounds bond with p-block elements that are in hypervalent states.[10] In contrast, many metals close to the s-d border such as Ca, Sc, Y, La, Ce, Th incline to form superhydrides with exceedingly high H compositions in which all H atoms form extended lattices (Fig. 1a-c).[11–18] The H lattices in these compounds feature strong covalent H-H bonding, high electron density and strong electron-phonon coupling, giving rise to superconductivity at very high temperatures.[19–21]

A large chemical driving force is needed to explain the formation of metal superhydrides. The available assumptions focus either at the volume effect or charge transfer between metals and hydrogen,[7,11] which can be examined by splitting the enthalpy changes into the changes of internal energy and the PV term (Fig. 1d-f and Supplementary Fig. 1). Generally, metallic hydrogen phases such as Cs-IV[22,23] show significant volume reduction while comparing with molecular phases such as $C_2/c$[24,25], but their internal energies are much higher at lower pressures. The volume reductions are even more significant for many superhydrides. However, for most of the superhydrides, $\Delta PV$ increases with pressure. The volume of H (as defined in methods section) in $LaH_{10}$ becomes higher than Cs-IV H at about 170 GPa and even higher than molecular $C_2/c$ phase at about 330 GPa (Fig. 1f). Therefore, it cannot be a consistent driving force for the formation of superhydrides. The volume reduction of superhydrides at lower pressures is more of a result of strong chemical interactions with high symmetry constrains.

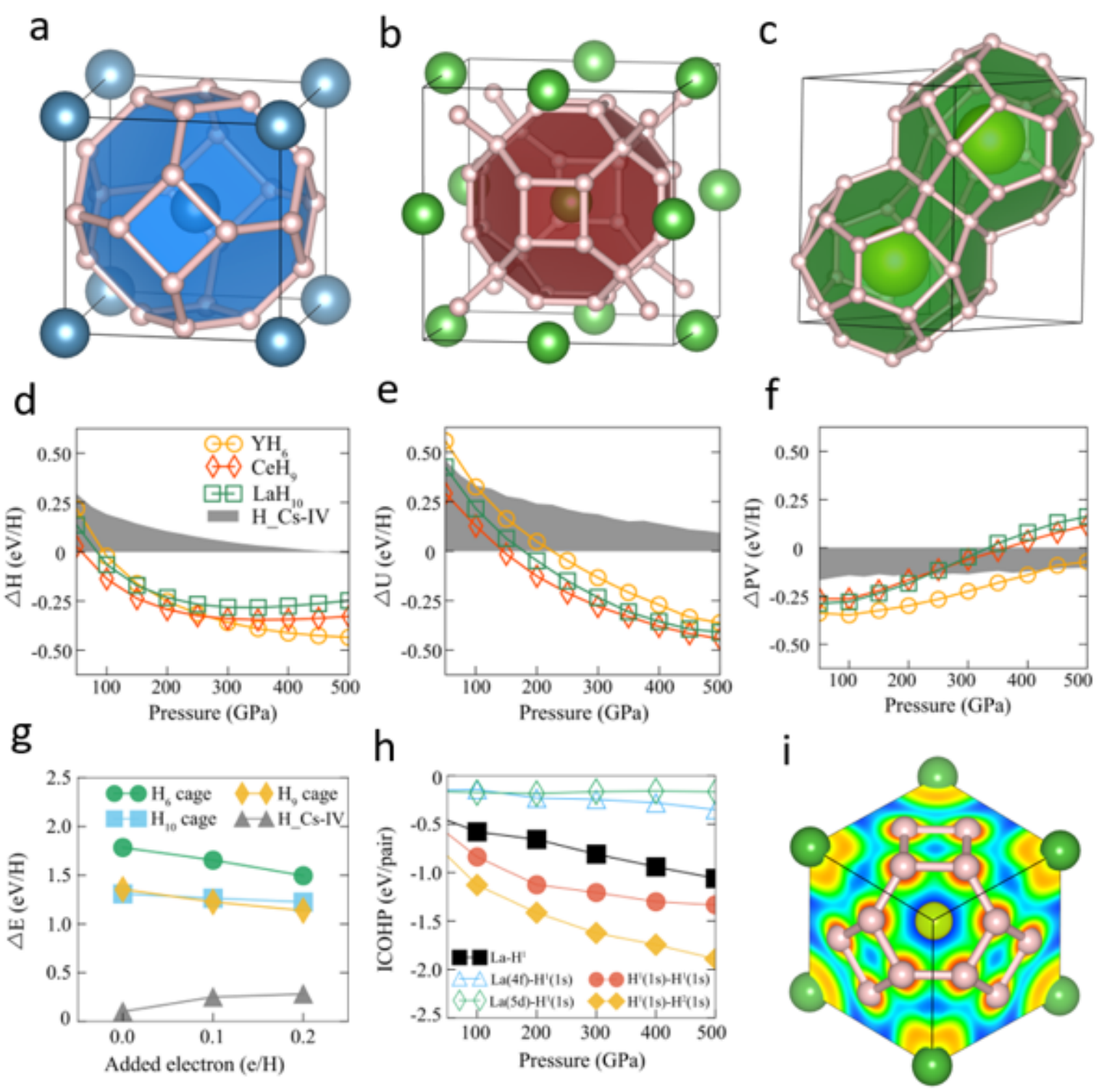

**Fig. 1 | Structure, stability and bonding features of metal superhydrides. a. – c.** The sodalite structures of three major types of metal superhydrides, including $CaH_6$ in $Im\bar{3}m$ structure, $LaH_{10}$ in $Fm\bar{3}m$ structure, and $CeH_9$ in $P6_3/mmc$ structure. The large blue, green and yellow balls represent the Ca, La, and Ce atoms; and the small white balls represent the H atoms. **d. – f.** The extracted enthalpy ($\Delta H$), internal energy ($\Delta U$) and $\Delta PV$ term of hydrogen in superhydrides relative to pristine hydrogen in molecular phase as functions of pressure (see Methods). The shaded areas show $\Delta H$, $\Delta U$, and $\Delta PV$ of metallic H in Cs-IV structure. **g.** The internal energies of hydrogen lattices (after removing metal atoms from superhydridies) in $CaH_6$, $LaH_{10}$, and $CeH_9$ and of metallic hydrogen in Cs-IV structure relative to molecular hydrogen phase as functions of added charges. **h.** The integrated crystalline orbital Hamiltonian population (ICOHP) of La–H, La 5d – $H^1$ 1s, La 4f – $H^1$ 1s, $H^1$ – $H^1$, and $H^1$ – $H^2$ as functions of pressure. **i.** The electron localization function (ELF) of $YH_6$ at 300 GPa, viewed from (111) direction. A [111] cutoff plane is placed at a distance of 4.40 Å from the origin.

In contrast to PV, the relative internal energies of superhydrides decrease steadily while comparing with molecular and atomic hydrogen phases (Fig. 1e and Supplementary Fig. 1), indicating strong chemical interactions between H and metals. Electron doping theory attributes this chemical interaction to the charge doping from metals to H, which seemingly explains the destabilization of $H_2$ molecules because the extra electrons will occupy their anti-bonding states. However, calculations show that the excess charges also destabilize extended H lattices and the overall effect to H polymerization is insignificant (Fig. 1g). The energy of Cs-IV phase of metallic H actually goes up comparing with $H_2$ $C_2/c$ phase while adding electrons (Fig. 1g). Furthermore, the charge doping from metal to hydrogen in superhydrides decreases with increasing pressure, distinctly opposing the trend of internal energy. Besides charge doping, the strong covalent bonding between metals and hydrogen seems to be another candidate of the driving force. However, both integrated Crystalline Orbital Hamiltonian Population (ICOHP)[26] (Fig. 1h) and Electron Localization

Function (ELF)[27] (Fig. 1i) reveal that the M-H bond strengths are significantly weaker than H-H bonds and therefore can hardly cause the formation of extended H lattice. Especially, the very small ICOHP values between the 5d/4f orbitals of La and H 1s show that these orbitals do not play major roles in forming metal superhydrides.

A new set of questions emerged after putting together all the recently predicted and synthesized metal superhydrides. Several different structures are often found for superhydrides with same H compositions and their energy order strongly depends on metals. Although, the hydrogen lattices in these structures show intricate geometry, the metal atoms usually form simple lattices such as face-centered cubic (FCC), hexagonal close pack (HCP), simple hexagonal etc., or lattices slightly deformed from them. For example, most of the $MH_9$ except $PrH_9$ and $PaH_9$, adopt hexagonal structures in which the metal atoms form HCP or SH lattices. Also, the $MH_{12}$ structure with highest symmetry ($Fm\bar{3}m$) consists of FCC metal lattice and H cubo-octahedra locating at the octahedral sites. However, $BaH_{12}$ in this structure is significantly higher in energy than a distorted structure with $P2_1$ symmetry, although the Ba lattice in the latter structure forms a slightly distorted FCC lattice.[28] The stability trend of these structures is hard to be explained by the direct chemical interactions between metals and H.

In the following sections, we will show that instead of directly bonding with H, the presence of the metals significantly enhances the stability of the extended hydrogen lattices in superhydrides through a chemical template effect. The new theory can explain the formation of the superhydrides, the trend of metals that can form superhydrides across the periodic table, the preferences of superhydride structures for different metals, etc. It can also help to search new superhydrides, especially those with higher H compositions and with mixed metals.

## Results and discussions

**The electron localizations in metal lattices.** The first step toward the mechanism of superhydrides formation is noticing that the electron localizations are ubiquitous in metals including the metal sub-lattices of superhydrides. In high-pressure electrides (HPE) such as hP4 Na, electrons occupy the local quantum orbitals at the interstitial site and play the role of anions of ionic compounds.[29,30] This view can be extrapolated to many low-pressure structures of metals in which electrons only partially occupy the interstitial orbitals, rendering them in metallic states. For example, the La atoms in $LaH_{10}$ at 300 GPa correspond to a face-centered cubic (FCC) lattice at 12.4 GPa. Its ELF exhibits maxima with considerable values of 0.62 and 0.45 at the centers of octahedral ($E^O$) and tetrahedral ($E^T$) sites, suggesting some crystal orbitals have large distributions at these areas (Fig. 2a). As a matter of fact, the three highest occupied crystal orbitals (CO) at the Γ point that are 2.96 (HOMO-2), 2.38 (HOMO-1) and 2.33 (HOMO) eV below $E_F$, show maxima and large contributions at $E^O$ sites, both $E^O$ and $E^T$ sites, and $E^T$ sites, respectively (Fig. 2b-d). On the other hand, the charge distribution of FCC La shows no maxima at $E^O$ and $E^T$ sites; therefore, it is not an HPE. Similar localizations also happen to other metal lattices, for example, Sc BCC lattice in $ScH_6$ (Fig. 2e) and Y HCP lattice in $YH_9$ (Fig. 2f).

The strengths and the sites of the electron localizations strongly depend on the metal atoms and the geometry, and they evolve systematically with the size of the metal lattices. While lattice constant reduces, the general trend of the electron localization is from sites with less to sites with more surrounding atoms (Fig. 2g-h). For example, while the unit length of an FCC Ca lattice reduces, the electron localization shifts from sites between two atoms (bond center), to $E^T$ sites, and then to $E^O$ sites (Fig. 2i-j). For metals in the same period, the electron localization decreases with increasing atomic number. They are the strongest for s-block metals and early transition metals and become much weaker for late transition and p-block metals such as Al (Fig. 2k-n). Interestingly, the localizations for early transition metals such as Y, La, and Hf etc. are strong on both $E^O$ and $E^T$ sites, an important feature that will stabilize $MH_{10}$ superhydrides. It is important to notice the difference between the electron localizations and electron distributions at the interstitial sites. Metals like Al cannot stabilize H lattice and form superhydrides, although they have very high electron densities at the interstitial sites.

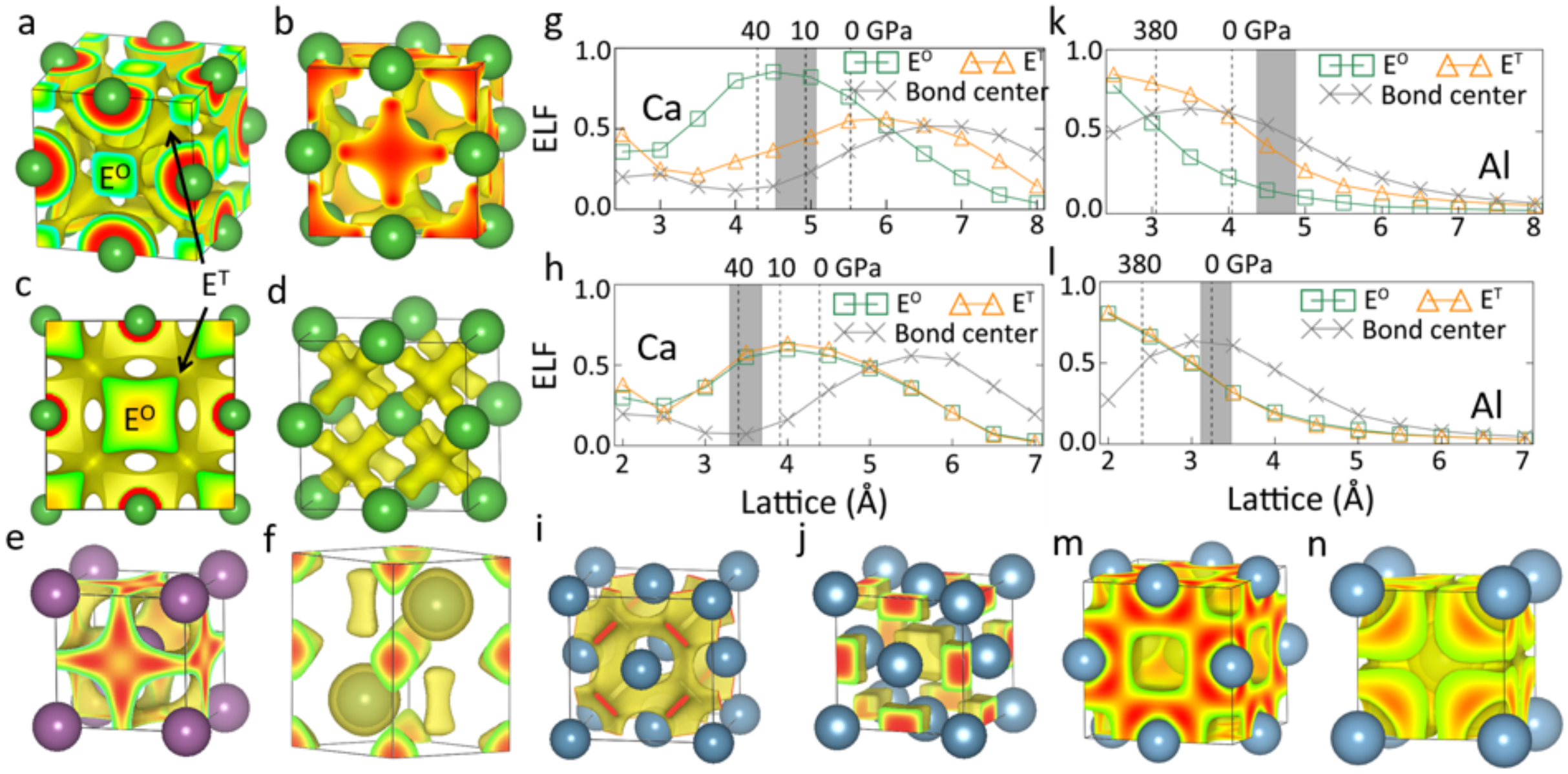

**Fig. 2 | The electron localizations and their evolutions in metal lattices. a.** The ELF of La sublattice in $LaH_{10}$ at 300 GPa. **b. – d.** Crystal orbitals HOMO, HOMO-1, and HOMO-2 at the Γ point of La sublattice in $LaH_{10}$. **e.** The ELF of Sc BCC lattice in $ScH_6$ at 100 GPa. **f.** The ELF of Y HCP lattice in $YH_9$ at 300 GPa. **g.** ELF values at interstitial sites in Ca FCC as functions of unit length. The shaded area shows the unit lengths of Ca lattice in a conceived $CaH_{10}$ at pressures from 100 to 300 GPa. **h.** ELF values at interstitial sites in Ca BCC as functions of unit length. The shaded area shows the unit lengths of Ca lattice in $CaH_6$ at pressures from 100 to 300 GPa. **i.** ELF of Ca FCC lattice with a unit length of 6.5 Å. At this length, FCC Ca show large electron localizations at $E^T$ sites. **j.** ELF of Ca FCC lattice with a unit length of 4.7 Å. This length corresponds to Ca FCC lattice in a conceived $CaH_{10}$ under 300 GPa. **k.** ELF values at interstitial sites in Al FCC as functions of unit length. The shaded area shows the unit lengths of Al lattice in a conceived $AlH_{10}$ at pressures from 100 to 300 GPa. **l.** ELF values at interstitial sites in Al BCC as functions of unit length. The shaded area shows the unit lengths of Al lattice in a conceived $AlH_6$ at pressures from 100 to 300 GPa. **m.** ELF of Al FCC lattice under ambient pressure. **n.** ELF of Al BCC lattice under ambient pressure. Both ELFs show that electrons in Al lattices localize mainly around Al atoms due to the occupation of Al orbitals.

**The building units of H lattices.** The second key feature of superhydrides is that the H lattices consist of unique and intrinsic building units (Fig. 3a) that are positioned right at the active regions of the metal templates. They can be identified by the geometry, symmetry and the crystal orbitals of H lattices (Fig. 3b, c). Some units are straightforward to identify by their appearance in the lattice, such as $H_6$ hexagons and $H_4$ squares in $MH_6$, and $H_8$ cubes and $H_5$ tetrahedrons in $MH_{10}$; whereas some others are quite unexpected. For example, the H units in $MH_9$ are not $H_5$ and $H_6$ rings, but rather a $H_6$ corona and a $H_8$ bipyramid (Fig. 3a). These two units share most of the symmetries of $MH_9$, and many occupied crystal states localize on them (Fig. 3c). The ways that H lattices are divided into building units are also corroborated by their energies that are calculated by use of a He-matrix model (see Methods). Among the 6 building units in $MH_6$, $MH_9$ and $MH_{10}$, the $H_6$ hexagon, the $H_8$ cube and the $H_6$ corona are significantly lower in energy. Also, their energies decrease for about 0.5 eV/atom while pressure increases from 100 GPa to 300 GPa (Fig. 3d).

The remarkable stability of $H_6$ hexagon, $H_8$ cube and $H_6$ corona originates from an important feature of H-H bond. Due to the quantum resonance, these bonds are conjugated and delocalized in the same way as the C-C 2p π bonds in organic molecules of which the stability is ruled by the aromaticity.[31,32] Because of the topology of the π bonds, the aromatic molecules need to assume a planar geometry and their electron counting needs to satisfy 4n+2 rule, which ensures a gap between the fully occupied and unoccupied orbitals. However, in contrast to C π bonds, the conjugation of H-H bonds is not constrained inside the same plane. The corresponding three-dimensional aromaticity depends on the symmetry and the number of H in the cluster, and the above three H clusters are all aromatic. The energy levels of $H_6$ hexagon resemble the energy levels of the benzene ring,[31] and the highest occupied (HOMO) and the lowest unoccupied molecular orbitals (LUMO) are doubly degenerate (Fig. 3e). In contrast, the HOMO and LUMO of $H_8$ cube are triply degenerate (Fig. 3f). The HOMO-LUMO gaps of $H_6$ hexagon, $H_8$ cube and $H_6$ corona are 8.50, 10.15 and 7.77 eV, respectively. Many other aromatic H clusters are also identified as building units in various superhydrides (Supplementary Section II).

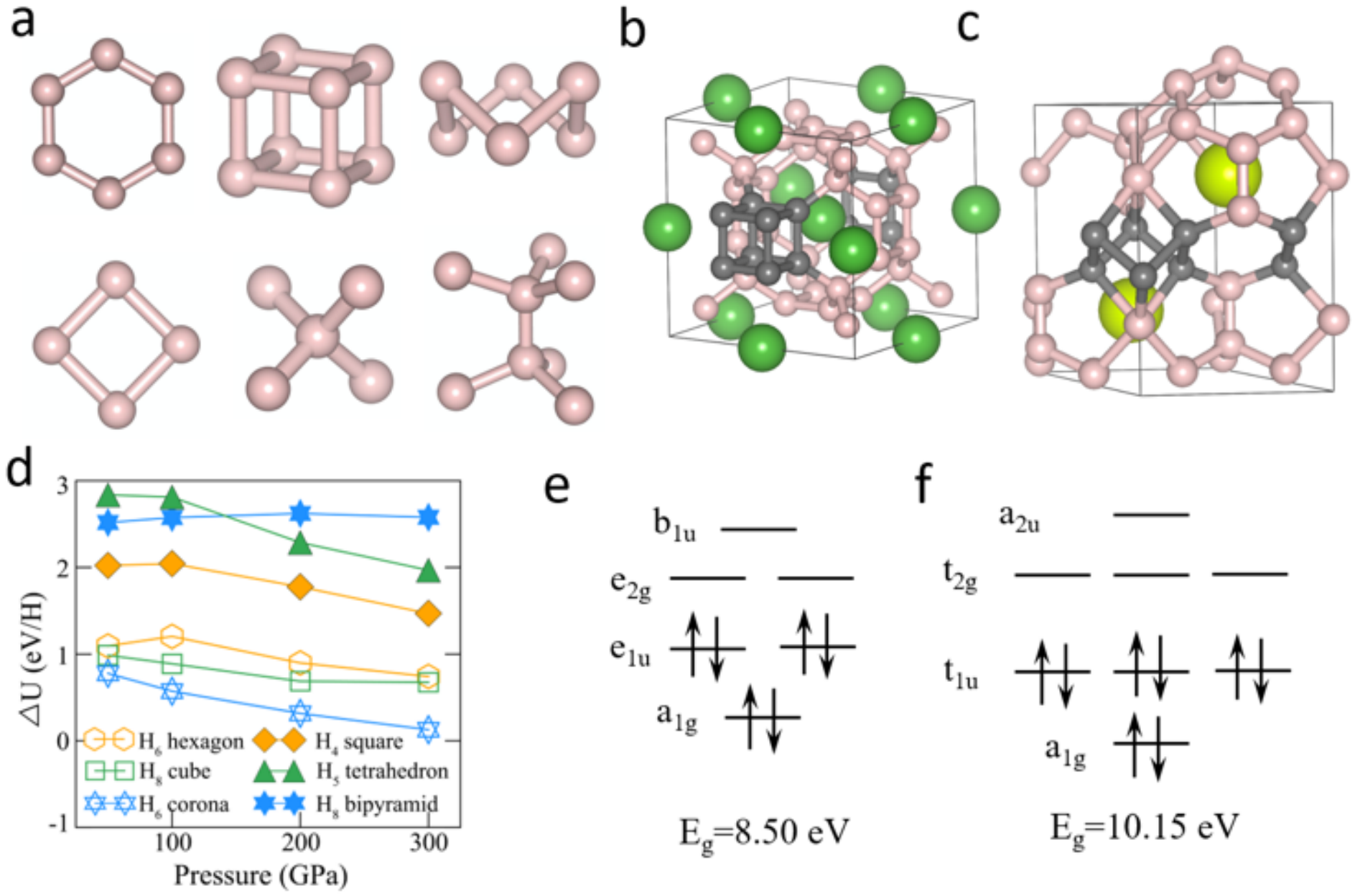

**Fig. 3 | The building units of H lattices in superhydrides. a.** The six building units of the H lattices in $MH_6$, $MH_{10}$ and $MH_9$ superhydrides, including the $H_6$ hexagon, the $H_4$ square, the $H_8$ cube, the $H_5$ tetrahedron, the $H_6$ corona and the $H_8$ bipyramid. **b.** The $H_8$ cube, a building unit of $H_{10}$ lattice in $LaH_{10}$. **c.** The $H_6$ corona, a building unit of $H_9$ lattice in $CeH_9$. **d.** The energies of the H units relative to $H_2$ molecules as functions of pressure. The calculations are performed by use of the He matrix model. **e.** The symmetries and the calculated energy levels of a $H_6$ hexagon. **f.** The symmetries and the calculated energy levels of a $H_8$ cube.

**The assembly of H covalent network on metal templates.** While the ELF and the COs of metal sublattices in superhydrides are overlaid on the corresponding H lattices, a striking feature of the superhydrides emerges unexpectedly. The H lattice matches excellently with the ELFs (Fig. 4a-c) and relative COs (Fig. 4d) of the metal lattices that include no information of H atoms at all. The $H_{10}$ lattice in $LaH_{10}$ consists of two types of H, $H^1$ that forms $H_8$ cubes locating at the $E^O$ sites, and $H^2$ locating at $E^T$ sites. Furthermore, some COs of the $H_{10}$ lattice also consist of orbitals locating at $E^O$ and $E^T$ sites. At Γ point, the lowest unoccupied CO (Fig. 4e) consists of orbitals locating at both $E^O$ and $E^T$. Therefore, while the La and the $H_{10}$ lattices are interposed together, the occupation of $E^O$ and $E^T$ orbitals in La valence bands will naturally dope $H_{10}$ by occupying its conduction band states. Indeed, the highest occupied state of $LaH_{10}$ at the Γ point consists of orbitals at $E^O$ and $E^T$ (Fig. 4f). Therefore, instead of direct charge transfer from La atoms to H atoms, large part of the electron density of La lattice already localizes around the consisting motifs of H lattices, forming a chemical template awaiting and assisting the assembly of the H lattice.

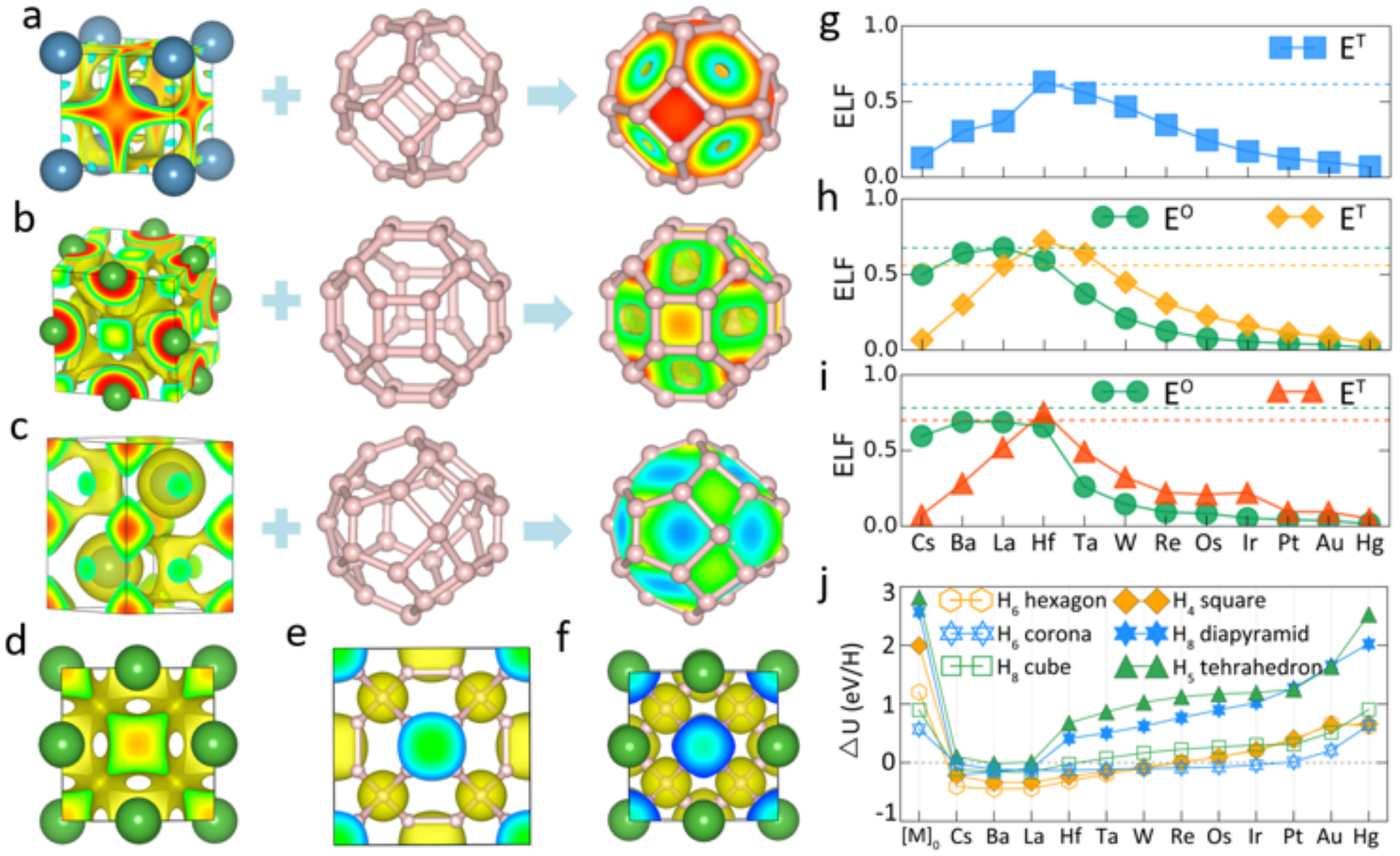

**Fig. 4 | Assemble H lattices in metal templates. a.** The ELF of a Ca BCC lattice overlaid on $H_6$ lattice. **b.** The ELF of La lattice in $LaH_{10}$ at 300 GPa, overlaid on the $H_{10}$ lattice. **c.** The ELF of Ce lattice in $CeH_9$ at 300 GPa, overlaid on the $H_9$ la ttice. **d**. The density distribution of state HOMO-1 at Γ point of La lattice in $LaH_{10}$ at 300 GPa. **e.** The density distribution of the lowest unoccupied state (LUMO) at Γ point of the $H_{10}$ lattice (after removing La atoms) in $LaH_{10}$ at 300 GPa. **f.** The density distribution of the highest occupied state (HOMO) of $LaH_{10}$ at Γ point at 300 GPa. **g.** The ELF values at $E^T$ sites of 6$^{th}$ row metal BCC lattice with a unit length corresponds to $MH_6$ at 100 GPa. The dashed line corresponds to the value at $E^T$ site of Ca BCC lattice in $CaH_6$. **h.** The ELF values at $E^O$ and $E^T$ sites of 6$^{th}$ row metal FCC lattice with a unit length corresponds to $MH_{10}$ at 100 GPa. The dashed lines correspond to the values at $E^O$ and $E^T$ sites of La FCC lattice in $LaH_{10}$. **i.** The ELF values at $E^O$ and $E^T$ sites of 6$^{th}$ row metal HCP lattice with a unit length corresponds to $MH_9$ at 100 GPa. The dashed lines correspond to the values at $E^O$ and $E^T$ sites of Ce HCP lattice in $CeH_9$. **j.** The energies of the H units relative to $H_2$ with the presence of metals in the 6$^{th}$ row of periodic table, calculated by use of the He matrix at 100 GPa.

The close match of the metal ELF and the H lattice is found for all metal superhydrides despite the large variation of structures and symmetries, such as $MH_6$ in $Im\bar{3}m$ structure (M= Sc, Ca, Y), $MH_9$ in $P6_3/mmc$ structure (M=Ce, Pr, La, etc.), $MH_{10}$ in $P6_3/mmc$ structure (M=Y, Ac, etc), $MH_{12}$ in $Fm\bar{3}m$ structure (M= Ba) and $MH_{16}$ in *P6/mmm* structure (M=La, etc.) and many others (Fig. 4g-i and Supplementary Section III). The metals in these superhydrides may adopt many different structures, including FCC, body-center cubic (BCC), hexagonal close-packed (HCP), simple cubic (SC), simple hexagonal (SH), or structures that deformed along high symmetry directions of them.

These sub-lattices are not necessary the stable structures of metals. Especially, for non-cubic superhydrides, the stresses of the metal lattices are usually not hydrostatic (Supplementary Table 3). For example, the normal stresses of the Ce lattice in $CeH_9$ at 300 GPa are $\sigma_1=\sigma_2=1.1$ GPa, $\sigma_3=15.1$ GPa. Nevertheless, all metal lattices show large electron distributions in the interstitial regions that match nicely with the locations and patterns of the H lattices in superhydrides, even if many superhydrides are in low symmetry structures, such as $R\bar{3}m$ $SrH_6$ and Immm $Ti_2H_{13}$ and the corresponding ELF and H lattices exhibit complicated geometry features. On the other hand, the strength of the template effect that can be estimated by the ELF values at the interstitial sites strongly depends on metals. While they are the strongest for metals at the s-d border, they decline quickly for metals away from that region and become insignificant for late transition metals (Supplementary Fig. 2). For example, in an FCC Ir lattice of a conceived $IrH_{10}$ compounds, the maxima of ELF are no longer at the interstitial sites but in the regions between neighboring Ir atoms.

The presence of the metal sub-lattices greatly improves the stability of H building units due to the chemical template effects and the strength of template as measured by ELF intensities strongly depend on the metal (Fig. 4g-i) as well as the pressure (unit length). We employ high throughput calculations using the He matrix model that allows comparing energies of H clusters with $H_2$ molecules in the same chemical environment (see Methods). The results show that the presence of the metal atoms can lower the energies of H units relative to $H_2$ for about 0.5 to 3 eV, including non-aromatic ones such as $H_4$ squares and $H_5$ tetrahedrons (Fig. 4j and Supplementary Fig. 3). This template effect strongly depends on metals. The most profound changes happen to the elements around the s-d border and decline with increasing number of d electrons, which is consistent with the trend of ELF (Fig. 4g-i and Supplementary Fig. 2). Among row 6 elements, those sit at the s-d border such as Ba and La can lower the energies of all H units except $H_5$ below $H_2$ at 300 GPa; whereas late transition metals such as Pt and Au show much weaker effect. Very few elements, including Ba, La and Th etc, could bring the energy of $H_8$ cube below $H_2$.

Pressure is essential to the stability of H lattice, by virtue of directly lowering the energies of H units and influencing the effect of metal templates. The distances between metal atoms are larger under lower pressure, reducing the electron density in the interstitial region and weakening the chemical driving force from the templates. The expansion of the metal lattices and the reduction of the template effect will also happen while more H atoms are packed into them, therefore placing limits to H compositions of superhydrides.

**The chemical template theory.** The conventional theory of chemical bonds in molecules and solid compounds take free atoms as the initial states.[33] The charge transfer and the bond energies are obtained by referring to free atoms and their quantum orbitals. While describing superhydrides $MH_n$, it is advantageous to take the sub-lattices, including metal and anion ($H_n$) lattices respectively, as the starting points. The Hamiltonian of compounds consisting of metals and anions can be written as $H = H_M + H_A$, in which $H_M$ and $H_A$ are the Hamiltonians of the metal and the anion sublattices. For ionic and largely polarized compounds, the valence states that are important for binding energy are more significantly determined by $H_A$, giving ground for treating the effects of metal lattices as a perturbation. Indeed, in highly ionic compounds, the valence states $\psi_0$ mainly consist of the local orbitals around the anions.

While using stationary perturbation theory and treating $H_M$ as the perturbation, the energy change of the valence state $\psi_0$ due to the presence of the metal sublattice is

$$\Delta\varepsilon = \langle\psi_0|H_M|\psi_0\rangle. \quad (1)$$

Assuming $\psi_M^i$ are the crystal orbitals of the metal sublattice, $H_M$ can be expressed as

$$H_M = \sum_i \varepsilon_M^i |\psi_M^i\rangle\langle\psi_M^i|, \quad (2)$$

and consequently,

$$\Delta\varepsilon = \sum_i \varepsilon_M^i \langle\psi_0|\psi_M^i\rangle^2. \quad (3)$$

If we assume only one crystal orbital of the metal sublattice $\psi_M^1$ has large overlap with $\psi_0$, $\Delta\varepsilon \approx \varepsilon_M^1\langle\psi_0|\psi_M^1\rangle^2$. Because $\varepsilon_M^i < 0$, the large overlap integral $S_{0\mathrm{M}} = \langle\psi_0|\psi_M^1\rangle$ will lower $\Delta\varepsilon$ and the valence state energy, and therefore will stabilizes the compound. Therefore, the presence of the chemical template could significantly stabilize the compounds because it optimizes the electron distribution in both anion and cation sublattices.

The concept of chemical template provides a very different view to chemical bonds in solid compounds. The strong chemical interactions due to the template effect do not associate with large electron relocations, in contrast to both ionic and covalent bonds. The electron distributions in superhydrides are optimal not only to the whole compound but also to its consisting metal and H sub-lattices, which maximize the stability. Taking $LaH_{10}$ as an example, the summation of the density distributions of the sub-lattices $\rho(La+H_{10}) = \rho[La\ lattice] + \rho[H_{10}]$ resemble very nicely $\rho(LaH_{10})$ (Supplementary Fig. 4a), which can be seen more clearly by the fact that $\Delta\rho = \rho(LaH_{10}) - \rho(La+H_{10})$ is quite small. As a matter of fact, the integrated transferred charge calculated from $\Delta\rho$ is about -0.15e for La (Supplementary Fig. 4b), which is about 10 times smaller than the Bader charge calculated from $\rho(LaH_{10})$ (Supplementary Fig. 4c). The negative value indicates an electron transfer to La, which can be misinterpreted as the presence of anionic La,[34] while it is actually the transfer of a small portion of charges around $E^O$ and $E^T$ back to La.

**Chemical templates in single-metal superhydrides.** Chemical template theory not only can explain the formation of superhydrides but also can explain the energy order of different structures and guide the search of new superhydrides. After intense search of new superhydride compounds and structures in the past several years, it becomes clear that almost every $MH_n$ composition corresponds to several structures with very different space groups. There is lack of a convincing and all-embracing theory explaining why different metals prefer certain structures. Many intricate structure preferences of superhydrides can be understood in the framework of chemical template theory. For example, between the two $MH_9$ structures[13,18,35], $P6_3/mmc$ ($CeH_9$) and $F\bar{4}3m$ ($PrH_9$), most metals prefer the former except $PrH_9$ and $PaH_9$ that are found in $F\bar{4}3m$ structure (Fig. 5a). In contrast to metals like Cs and Ba that show no significant electron localizations at $E^O$ sites and to metals like Y and La that show strong localizations bridging $E^T$-$E^T$ pairs (Fig. 5b), both features can lower the energy of $P6_3/mmc$ structure by relaxation in (111) direction, Pr and Pa show strong location at $E^O$ site but no strong pair-bridging (Fig. 5c).

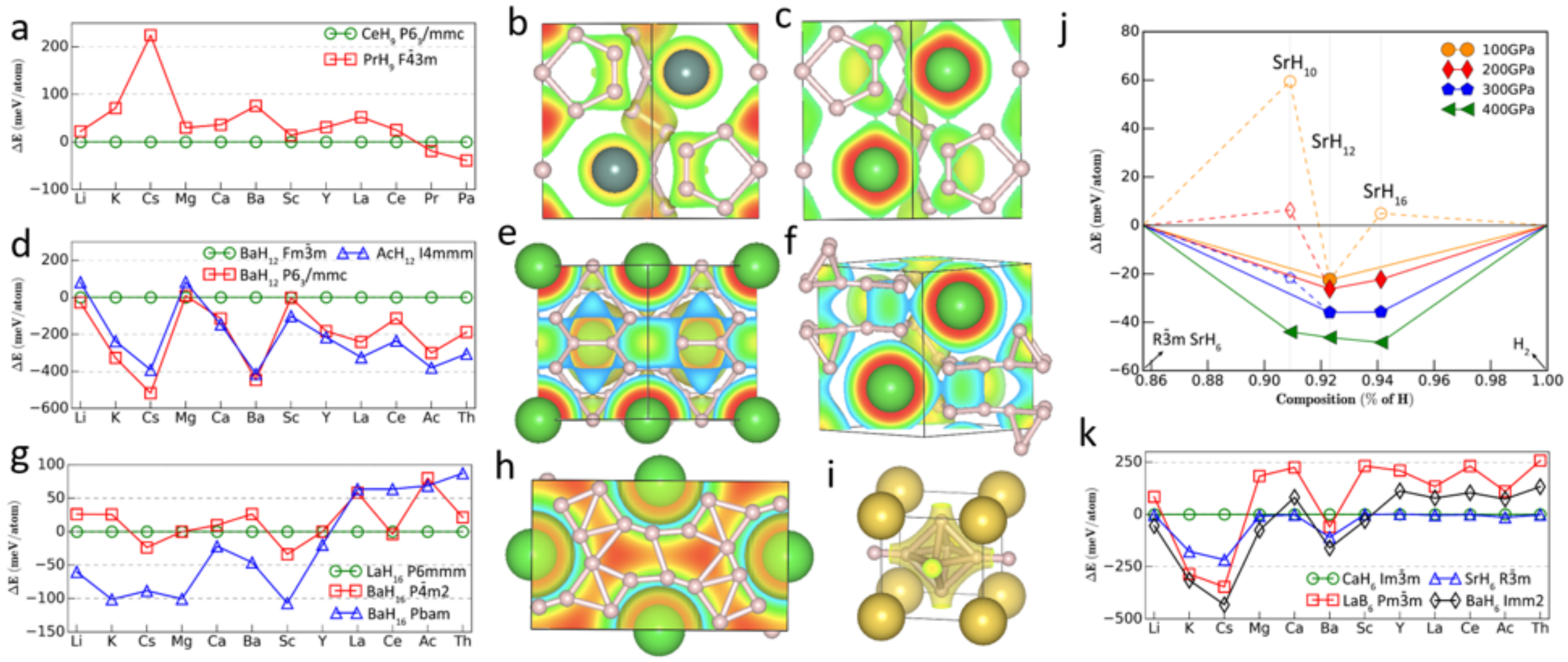

**Fig. 5. The chemical templates and the structures of single-metal superhydrides. a.** Structure preferences of $MH_9$ for selected metals. Two major structures, $P6_3/mmc$ and $F\bar{4}3m$ are compared. **b.** The ELF of Y sublattice in $YH_9$ in the $P6_3/mmc$ structure. **c.** The ELF of Pr sublattice in $PrH_9$ in the $P6_3/mmc$ structure. **d.** Structure preferences of $MH_{12}$ of selected metals. **e.** The ELF of Ba sublattice in $BaH_{12}$ in the $Fm\bar{3}m$ structure. **f.** The ELF of Ba sublattice in $BaH_{12}$ in the $P6_3/mmc$ structure. **g.** Structure preferences of $MH_{16}$ of selected metals. **h.** The ELF of Sr sublattice in $SrH_{16}$ in the $Pbam$ structure. **i.** The ELF of Na sublattice in $NaH_6$ in the $Pm\bar{3}m$ structure ($LaB_6$). **j.** Partial convex hall of Sr-H superhydrides. $SrH_6$ is assumed to be stable throughout the pressure range. **k.** Structure preferences of $MH_6$ of selected metals.

In another example, the structure of $BaH_{12}$ has been thoroughly studied by DFT based structure search method in conjunction with an experimental work. Among the predicted cubic structures that matches the X-ray diffraction pattern, the $Fm\bar{3}m$ $BaH_{12}$ consists of highly symmetric H cubo-octahedra that occupy the $E^O$ sites of FCC Ba. However, the energy of this structure is significantly higher than the distorted cubic structures such as the $P2_1$ and the $Cmc2_1$ structures[28] (Fig. 5d). The cause of this symmetry reduction is due to the less significant and yet considerable electron localization at the $E^T$ sites that can be utilized to stabilize the H lattice in the structures with reduced symmetry (Fig. 5e). On the other hand, the HCP Ba lattice contains paired $E^T$ sites and allow lower symmetry of the polyhedrons. The search of $BaH_{12}$ superhydride structures by adding H atoms in the Ba HCP lattice found a $P6_3/mmc$ $BaH_{12}$ structure (Fig. 5f) that is only 25 and 16 meV higher in energy than $P2_1$ $BaH_{12}$ at 100 and 200 GPa, respectively. In case of several other metals, such as K and Cs, $P6_3/mmc$ is the most stable one among all known $MH_{12}$ structures.

$LaH_{16}$ is another promising structure with high H composition. La forms SH lattice in this structure and its ELF shows large electron localization at the prism interstitial sites (Fig. 5h). However, the ELFs of many other metals are distinctly different to La. Especially, there is lack of electron localization close to the metal hexagonal planes that are necessary to stabilize the $LaH_{16}$ structure. We performed a crystal structure search for $BaH_{16}$ and $ScH_{16}$ by adding H atoms to Ba and Sc SH and SC lattices. Two $MH_{16}$ structures are found including a new Pbam structure containing an SH metal lattice, and a $P\bar{4}m2$ structure containing a slightly deformed SC metal lattice. The latter structure has been found in a structure prediction study of $AcH_n$ compounds under high pressure.[36] Although the identified Pbam structure is lowest in energy for $BaH_{16}$, it is slightly above convex hull at pressures below 300 GPa, mainly due to the exceedingly stable $BaH_{12}$ compound. However, after examining all other metals with strong template effect, we found that $SrH_{16}$ in the same structure become stable at pressures above 186 GPa (Fig. 5j). $HfH_{16}$ is also stable, but only at pressures below 110 GPa.

A large-scale structure search based on chemical templates might help us find more stable superhydrides. For example, the strong electron localization at the body center of the simple cubic lattice formed by alkali metals and Ba suggests possible superhydrides in $LaB_6$ structure. As a matter of fact, the $MH_6$ superhydrides of most alkali metals and Ba are more stable in $LaB_6$ structure than $CaH_6$ structure (Fig. 5k). However, in most of the cases except $NaH_6$, their energies are slightly higher than $BaH_6$ Imm2 structure[37]. Among all the known structures, $NaH_6$ is most stable in $LaB_6$ structure (Fig. 5i). While constructing the convex hull of Na-H binary compounds, $NaH_6$ is found to be 22 meV above the convex hull at 200 GPa.

**Chemical templates in mixed-metal superhydrides.** While almost all possible binary superhydrides have been tried out, the hope of achieving higher Tc at lower pressure rely on the search of novel ternary superhydrides composed of two different metals.[12,38–41] Constructing a complete phase diagram of ternary superhydrides based on full-scale DFT calculations is extremely difficult, and the search of optimal compounds across the entire periodic table as we have done for binary superhydrides is an impossible task. To date, the diagrams of very few ternary superhydrides such as Li-Mg-H have been thoroughly searched, which predicted a metastable superhydride, $Li_2MgH_{16}$, that shows a $T_c$ of ~473K at 250 GPa.[38] The template theory allows us to assess the formation of mixed-metal superhydrides by studying only the metal lattices. High-throughput calculations show strong correlation between the enhancements of the template strength while mixing metals and the energy of formation of mixed metal superhydrides.

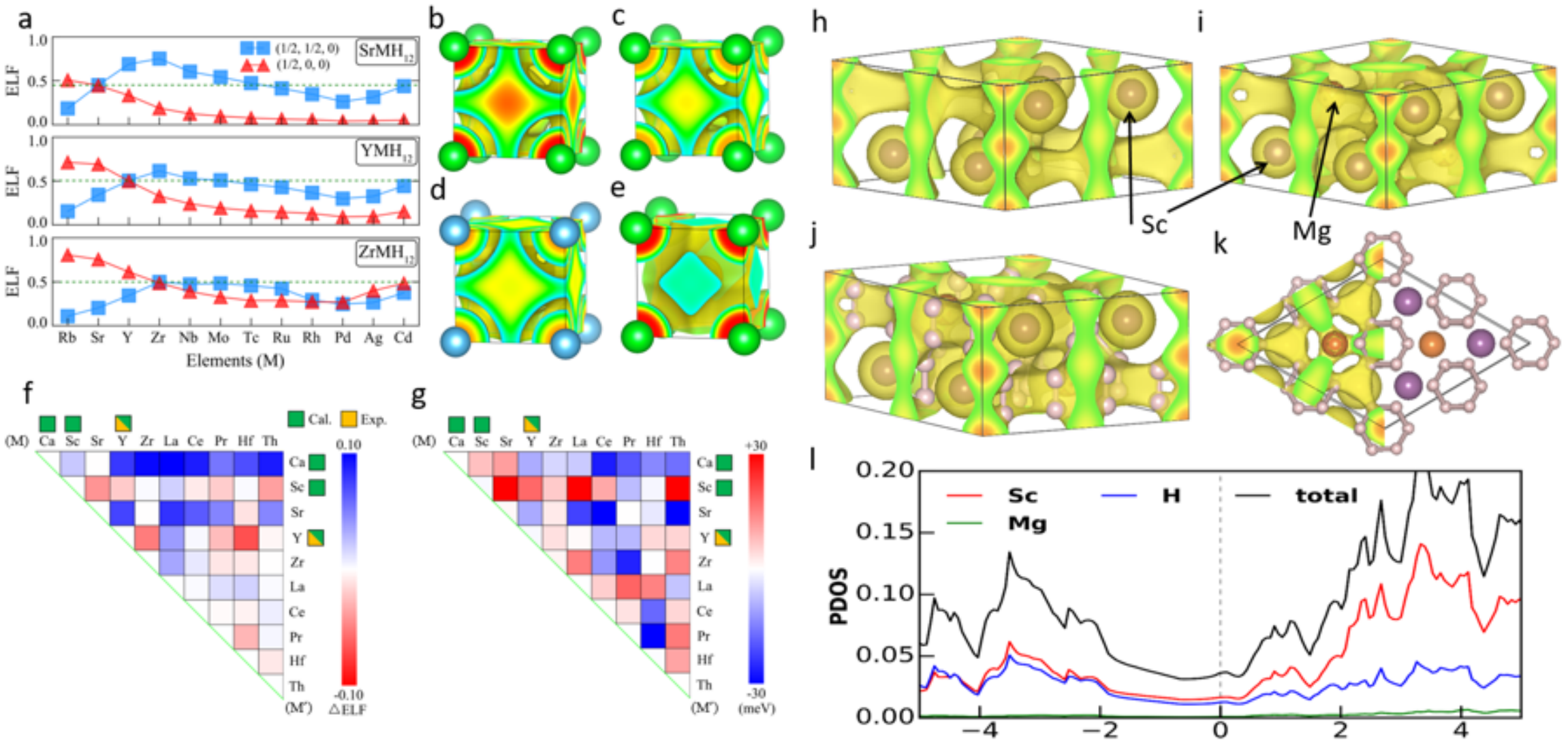

**Fig. 6 | The chemical template effects in mixed-metal superhydrides. a.** The ELF values of $MM'H_{12}$ mixed metal superhydrides at the interstitial points, including (1/2, 1/2, 0) and (1/2, 0, 0). M is Sr, Y and Zr, and M′ is a $5^{th}$ row metal element. **b. – e.** The ELF of $SrYH_{12}$, $SrZrH_{12}$,

$YZrH_{12}$ and $SrAgH_{12}$. **f.** The average ELF of $MM'H_{12}$ superhydrides at 100 GPa, $\Delta ELF = ELF^{1/2,1/2,0}_{MM'H_{12}} + ELF^{1/2,0,0}_{MM'H_{12}} - ELF_{MH_6} - ELF_{M'H_6}$, in which M and M′ are Ca, Sc, Sr, Y, Zr, La, Ce, Pr, Hf and Th. **g.** The reaction enthalpy of forming $MM'H_{12}$, $\Delta H = (H(MM'H_{12}) - H(MH_6) - H(M'H_6))/14$, at 100 GPa. In both (**f**) and (**g**) the squats filled with green and orange colors show $MH_6$ superhydrides that are predicted by DFT calculations and synthesized by DAC experiments, respectively. **h.** ELF of Sc sublattice in $Sc_3MgH_{24}$ superhydrides. **i.** ELF of $Sc_3Mg$ sublattice in $Sc_3MgH_{24}$ superhydrides. **j.** Side view of the ELF of $Sc_3MgH_{24}$ superhydrides. **k.** Top view of the ELF of $Sc_3MgH_{24}$ superhydrides. **l.** The total and projected density-of-states (DOS) of $Sc_3MgH_{24}$ superhydrides.

High throughput calculations based on the chemical template theory reveal two mechanisms of combining two metals to generate stable ternary superhydrides. In the first case, the combination of the two "template-active" metals might strengthen the effect. The ELF of two-metal lattices adapted from metal structures in $MH_6$, $MH_{10}$, and $MH_9$ at 100 GPa by partially replacing metal atoms (Fig. 6a-e and Supplementary Fig. 5) show the mixture with later transition metals greatly lowers the ELF values at the interstitial sites. In contrast, if both metals are close to s-d border such as Sr, Y and Zr (Fig. 6b-d), the resulting electron localization might be enhanced, which can be measured by $\Delta ELF = ELF^{1/2,1/2,0}_{MM'H_{12}} + ELF^{1/2,0,0}_{MM'H_{12}} - ELF_{MH_6} - ELF_{M'H_6}$ (Fig. 6f). More importantly, $\Delta ELF$ shows strong correlation with the stability of the ternary superhydrides that is calculated as the reaction enthalpies $\Delta H = [H(MM'H_{12}) - H(MH_6) - H(M'H_6)]/14$ (Fig. 6g). For example, while mixing Sr and Y, the average ELF increases (Fig. 6b) and the $SrYH_{12}$ is stable against the decomposition into $SrH_6$ and $YH_6$ (Fig. 6g). In contrast, the mixture of Y and Zr leads to lower average ELF (Fig. 6c) and correspondingly $YZrH_{12}$ is not stable (Fig. 6g). Similar trend can be found for $MH_9$ and $MH_{10}$ related two-metal superhydrides (Supplementary Fig. 6). A rough estimation of $T_c$ using the ratio of the density of hydrogen related states versus the total density of states[42] at the Fermi level show that the mixing metals in superhydrides can potentially improve $T_c$ (Supplementary Fig. 7).

In the second case, an active metal such as Li, Na and Mg etc. is mixed with a metal at the s-d border and enhance their template effect by doping electrons. For example, $Sc_3Mg$ in $P6_3/mmc$ structure show strong enhancement of the Sc template by adding Mg. The ELFs of the metal lattice with and without Mg show the same topology, but the values are significantly higher while adding Mg (Fig. 6h-i), which is caused by the electron transfer from Mg to Sc crystal orbitals. We thus conducted a structure search by adding H atoms into the $Sc_3Mg$ metal lattice, which leads to the discovery of a superhydride $Sc_3MgH_{24}$ also in $P6_3/mmc$ space group (Fig. 6j-k). This compound is stable against the decomposition into $ScH_6$, $MgH_4$ and $H_2$, with -0.54 meV/atom reaction enthalpy at 200 GPa. It shows considerable DOS at the Fermi level and might be a candidate superconductor under pressure (Fig. 6l). The results show that a large-scale study of the mixed metal lattices adapted from known intermetallic compounds and the change of their ELF is a promising and affordable approach to predicting mixed metal superhydrides in a massive scale. The mixed metal lattices adapted from $MH_6$, $MH_{10}$ and $MH_9$ in this work are also structures of known intermetallic compounds; and the Li-Mg lattice in $Li_2MgH_{16}$ is isostructural to Laves phase $MgCu_{12}$.[38]

## Conclusions

By studying the mechanism of metal superhydrides formation, we revealed a significant driving force of forming solid compounds that is not known before, *i.e.* the chemical template effect. Different to the traditional chemical bond theory that compares the compounds with free atoms, we view the formation of superhydrides as the interposition of metal and H sub-lattices. Our calculations showed large electron localizations at the interstitial sites due to the occupation of the crystal orbitals of the metal sublattices, forming chemical templates. They assist the dissociation of $H_2$ molecules and the formation of H covalent networks in superhydrides. Furthermore, the H sublattices consist of H building units that are aromatic despite that they are not planar. Calculations on a He-matrix model show that the presence of the metal sub-lattices can largely stabilize the building units of the H lattices. Furthermore, the chemical template theory explains the large structural variations and their energy orders for superhydrides with the same H composition. The potential of the chemical template mechanism in searching novel superhydrides especially with higher H compositions and mixed-metal are demonstrated. High-throughput calculations revealed a strong correlation between the strength of the chemical templates and the stability of the mixed metal superhydrides, indicating it can be used for a large-scale search of ternary and quaternary superhydrides. It will greatly enhance the efficiency of searching superhydride materials that might become superconducting at higher temperature and lower pressure.

## Methods

Solid-state density functional calculations. The underlying first-principles density functional theory (DFT) calculations were carried out by using the plane-wave pseudopotential method as implemented in Vienna *ab initio* Simulation Package (VASP).[43,44] The electron-ion interactions were described by the projector augmented wave pseudopotentials[45,46] and the used valence electrons are listed in Table 1. Some calculations of late transition metals, such as the ELF trend of various lattices under compression are done using pseudopotentials without including the outer-core p orbitals in the valence. Our test calculations show that the differences caused by this choice of psedupotentials is negligible. We used the generalized gradient approximation formulated by Perdew, Burke, and Ernzerhof[47] as exchange-correlation functional. A kinetic energy cutoff of 520 eV was adopted for wave-function expansion. The *k*-point meshes with interval smaller than $2\pi \times 0.03$ Å$^{-1}$ for electronic Brillouin zone to ensure that all enthalpy calculations converged within 0.02 eV/atom. The high-throughput first-principles calculations were performed by using the Jilin Artificial-intelligence aided Materials-design Integrated Package (JAMIP), which is an open-source artificial-intelligence-aided data-driven infrastructure designed purposely for computational materials informatics.[48]

| | | | | | | | | | | | |
|---|---|---|---|---|---|---|---|---|---|---|---|
| H | | | | | | | | | | | He |
| $1s^1$ | | | | | | | | | | | $1s^2$ |
| K | Ca | Sc | Ti | V | Cr | Mn | Fe | Co | Ni | Cu | Zn |
| $3s^2 3p^6 4s^1$ | $3p^6 4s^2$ | $3s^2 3p^6 3d^1 4s^2$ | $3p^6 3d^2 4s^2$ | $3p^6 3d^3 4s^2$ | $3p^6 3d^5 4s^1$ | $3p^6 3d^5 4s^2$ | $3p^6 3d^6 4s^2$ | $3p^6 3d^7 4s^2$ | $3p^6 3d^8 4s^2$ | $3p^6 3d^{10} 4s^1$ | $3p^6 3d^{10} 4s^2$ |
| Rb | Sr | Y | Zr | Nb | Mo | Tc | Ru | Rh | Pd | Ag | Cd |
| $4s^2 4p^6 5s^1$ | $4s^2 4p^6 5s^2$ | $4s^2 4p^6 4d^1 5s^2$ | $4s^2 4p^6 4d^2 5s^2$ | $4s^2 4p^6 4d^3 5s^2$ | $4p^6 4d^5 5s^1$ | $4p^6 4d^5 5s^2$ | $4p^6 4d^6 5s^2$ | $4p^6 4d^7 5s^2$ | $4p^6 4d^8 5s^2$ | $4p^6 4d^{10} 5s^1$ | $4d^{10} 5s^2$ |
| Cs | Ba | La | Hf | Ta | W | Re | Os | Ir | Pt | Au | Hg |
| $5s^2 5p^6 6s^1$ | $5s^2 5p^6 6s^2$ | $5s^2 5p^6 5d^1 6s^2$ | $5p^6 5d^2 6s^2$ | $5p^6 5d^3 6s^2$ | $5p^6 5d^5 6s^1$ | $5p^6 5d^5 6s^2$ | $5p^6 5d^6 6s^2$ | $5d^7 6s^2$ | $5p^6 5d^8 6s^2$ | $5d^{10} 6s^1$ | $5d^{10} 6s^2$ |
| | | Ce | Pr | Ac | Th | Pa | U | | | | |
| | | $5s^2 5p^6 4f^1 5d^1 6s^2$ | $5s^2 5p^6 4f^3 6s^2$ | $6s^2 6p^6 6d^1 7s^2$ | $6s^2 6p^6 6d^2 7s^2$ | $6s^2 6p^6 5f^2 6d^1 7s^2$ | $6s^2 6p^6 5f^3 6d^1 7s^2$ | | | | |

**Table 1.** The valence configurations of the pseudopotentials used in our solid-state DFT calculations.

**Energy analysis of H in superhydrides.** In order to compare the energy terms, including the enthalpies, the internal energies and the PV terms, of the hydrogen in superhydrides directly with pristine hydrogen phases under pressure, we deduct the corresponding energies of the metal hydrides with typical composition, and refer it to those of $C_2/c$ molecular $H_2$ phase. For example, the enthalpy $H$ of the H incorporated into $LaH_{10}$ beyond its typical composition $LaH_3$ is calculated as $\Delta H(\mathrm{H}) = [H(\mathrm{LaH_{10}}) - H(\mathrm{LaH_3}) - \frac{7}{2} \times \mathrm{H}(\mathrm{H}_2^{C2/c})]/7$. Similar formula is used to calculate $\Delta U(\mathrm{H})$ and $\Delta PV(\mathrm{H})$. These energy terms are used to compare with the corresponding values of pristine H phases, as shown in Fig. 1d-f and Supplementary Fig. 1.

**Electronic structure Analyses of solid compounds.** The electronic structures of metal superhydrides are calculated and analyzed by use of several methods, including the Bader's Quantum Theory of Atoms in Molecules (QTAIM),[49] the Electron Localization Function (ELF),[27] the Crystalline Orbital Hamiltonian Population (COHP) and integrated COHP (ICOHP),[26] and integrated differential charge density (see below), etc.

**Integrated differential electron density.** First, for a given metal superhydride ($MH_n$), three electron densities are calculated, including that of $MH_n$, of the metal lattice ($MH_n$ after removing all H), and of the H lattice ($MH_n$ after removing all metals). The differential electron density is then calculated as $\Delta\rho = \rho(MH_n) - \rho[\text{metal lattice}] - \rho[H_n \text{ lattice}]$. This differential density distribution is separated into regions by the surface(s) of $\Delta\rho=0$. Especially, all the metal atoms are surrounded by a surface of $\Delta\rho=0$ that is roughly spherical. In all the cases, $\Delta\rho>0$ inside the surface, indicating that electrons transfer toward metal atoms while interposing the metal and the H lattices. The total charge transfers to metals are calculated by integrating the differential electron density inside the $\Delta\rho=0$ surface. The charges shown in Supplementary Fig. 4c are negative because the electron charge is negative.

**Quantum chemistry calculations on H clusters.** The geometry parameters of all H clusters are adapted from the superhydrides optimized by solid-state DFT calculations at the studied pressures (100 and 300 GPa). Their molecular orbitals and the energy levels are calculated by using Gaussian 09 package.[50] The restricted open-shell B3LYP[51–53] and Hartree-Fock[54] methods are used for exchange and correlation functional for H cluster with even and odd number of H atoms respectively, respectively. The 6-31G(d) basis set is adopted for all the single-point calculations.

**He matrix model.** We build He matrix models to study the energies of various H clusters that are the building units of H lattices in superhydrides and compare them with $H_2$ molecules in the same chemical environment. Taking $LaH_{10}$ as an example, we first optimize its structure to a selected pressure, for example 300 GPa. A supercell is constructed by triple the units in all three directions. After that, we replace all the H atoms with He atoms. By keeping or removing La metal atoms, we constructed two models, $He_{320}$ and $La_{32}He_{320}$. The H clusters are created in these models by replacing corresponding He atoms with H atoms. For example, while modeling $H_8$ cubes, we replace 8 He atoms in a cube with H atoms. For comparison with $H_2$ molecules, we replace four pairs of He atoms in $H_{320}$ and $La_{32}He_{320}$ by four pairs of H atoms. In order to compare the energies of H cluster with $H_2$ molecules at the same chemical environment, each $H_2$ molecule is placed at the sites that are part of H cluster. Also, the $H_2$ molecules are positioned to maximize their distances so to minimize the factitious $H_2$-$H_2$ interactions in the model. The H coordinates in $H_2$ models are relaxed in order to maintain the lowest energy of $H_2$ molecules inside the He matrix. The relaxed H-H bond lengths in these models are close to that of $H_2$ molecules at ambient condition. The H atoms in the models of H clusters are not relaxed so that the H-H distances in these clusters are kept the same as in the H lattice of $LaH_{10}$ at the studied pressure. Similarly, the supercell He-matrix models, including $He_{96}$ and $Y_{16}H_{96}$, are constructed for $YH_6$, and the energies of $H_4$ square and $H_6$ hexagon are studied using these models. The supercell He-matrix models, including $He_{144}$ and $Ce_{16}He_{144}$, are constructed for $CeH_9$, and the energies of $H_6$ corona and $H_8$ bipyramid are studied using them. In high-throughput calculations, metals in superhydrides vary, and the corresponding He-matrix model are created from the $MH_n$ lattice optimized at the studied pressure.

**Stability of mixed-metal superhydrides.** The stability of the mixed-metal superhydrides is assessed by comparing their enthalpies with their constituent binary superhydrides. The enthalpy differences per atoms as shown in Supplementary Fig. 7 are calculated by the following formulae for three types of superhydrides based on $Im\bar{3}m$ $MH_6$, $P6_3/mmc$ $MH_9$ and $Fm\bar{3}m$ $MH_{10}$ superhydride structures. Specifically, the enthalpy differences are defined as:

$\Delta H = (H(\mathrm{MM'H_{12}}) - H(\mathrm{MH_6}) - H(\mathrm{M'H_6}))/14$, for $\mathrm{MM'H_{12}}$;

$\Delta H = (H(\mathrm{MM'H_{18}}) - H(\mathrm{MH_9}) - H(\mathrm{M'H_9}))/20$, for $\mathrm{MM'H_{18}}$;

$\Delta H = (H(\mathrm{M_nM'_mH_{40}}) - \frac{\mathrm{n}}{4} \times H(\mathrm{M_4H_{40}}) - \frac{\mathrm{m}}{4} \times (H(\mathrm{M'_4H_{40}}))/44$, for $\mathrm{M_1M'_3H_{40}}$, and $\mathrm{M_3M'_1H_{40}}$,

where $H(\mathrm{MM'H_{12}})$, $H(\mathrm{MH_6})$, $H(\mathrm{M'H_6})$, $H(\mathrm{MM'H_{18}})$, $H(\mathrm{MH_9})$, $H(\mathrm{M'H_9})$, $H(\mathrm{M_nM'_mH_{40}})$, $H(\mathrm{M_4H_{40}})$, and $H(\mathrm{M'_4H_{40}})$ represents the enthalpies of ternary and binary superdydrides. Their structures are optimized at 100 and 300 GPa. M and M′ represents the metals in 1-12 groups of 4-6 periods.

## Data availability

The authors declare that all of the data supporting the findings of this study are available within the paper and the Supplementary Information, and also from the corresponding authors upon reasonable request.

**Acknowledgements:** M.M. and Y.S. acknowledge the support of NSF CAREER award 1848141, M.M. also acknowledges the support of ACS PRF 59249-UNI6 and the support of California State University Research, Scholarship and Creative Activity (RSCA) award. We thank the computational resources provided by XSEDE (TG-DMR130005).

**Author contributions:** M.M. proposed the chemical template mechanism and designed the research. Y.S. conducted all the calculations. M.M. and Y.S. analyzed the results together. M.M. wrote the manuscript. M.M. and Y.S. plotted the figures together.

**Competing interests:** The authors declare no competing interests.

# Supplementary Information

## Chemical templates that assemble the metal superhydrides

Yuanhui Sun and Maosheng Miao*

Department of Chemistry and Biochemistry, California State University Northridge, CA USA

*Email: mmiao@csun.edu

# Supplementary Material

# Section I. Energetics of H incorporated in superhydrides

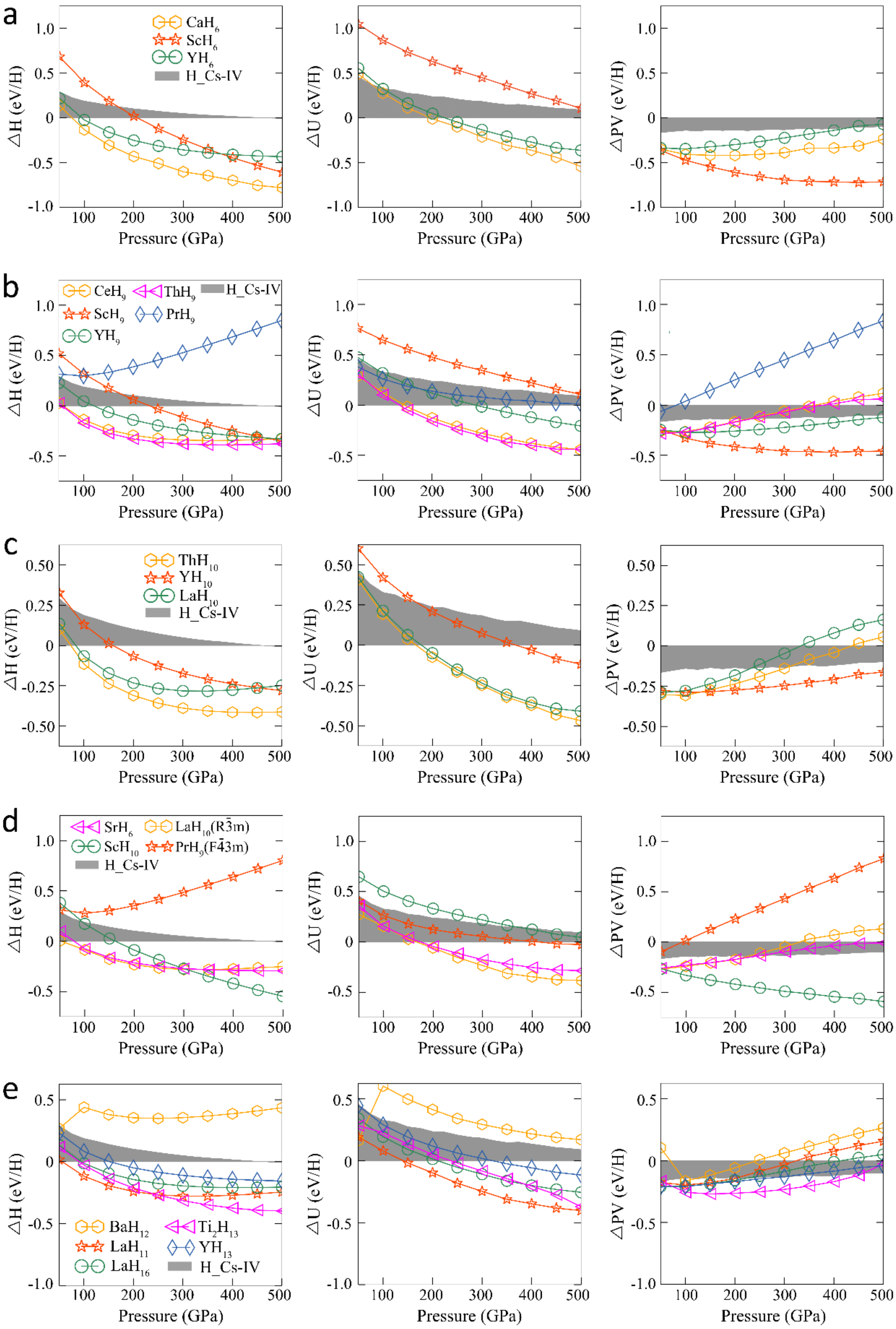

**Fig. 1:** The extracted enthalpy ($\Delta H$), internal energy ($\Delta U$) and $P\Delta V$ term of hydrogen in superhydrides relative to pristine hydrogen in molecular phase as functions of pressure. The investigated superhydrides include: **a**, $CaH_6$, $ScH_6$, $YH_6$ in $Im\bar{3}m$ structure; **b**, $CeH_9$, $ScH_9$, $YH_9$, $ThH_9$, $PrH_9$ in $P6_3/mmc$ structure; **c**, $ThH_{10}$, $YH_{10}$, $LaH_{10}$ in $Fm\bar{3}m$ structure; **d**, $SrH_6$ in $R\bar{3}m$ structure, $ScH_{10}$ in $P6_3/mmc$ structure, $PrH_9$ in $F\bar{4}3m$ structure, $LaH_{10}$ in $R\bar{3}m$ structure, and **e**, $BaH_{12}$ in $Fm\bar{3}m$ structure, $LaH_{11}$ in $P4nmm$ structure, $LaH_{16}$ in $P6/mmm$ structure, $Ti_2H_{13}$ in $Immm$ structure, $YH_{13}$ in $R\bar{3}m$ structure. All quantities are relative to that of molecular H $C_2/c$ structure. The shaded areas show $\Delta H$, $\Delta U$, and $P\Delta V$ of metallic H in Cs-IV structure.

## Section II. The building units of H lattices and their aromaticity

| Name | Compounds | Structure | Wavefunction | Energy Levels |
|---|---|---|---|---|
| $H_4$ tetrahedron | $MH_9$ M=Pr[1,2], Pa[3] | 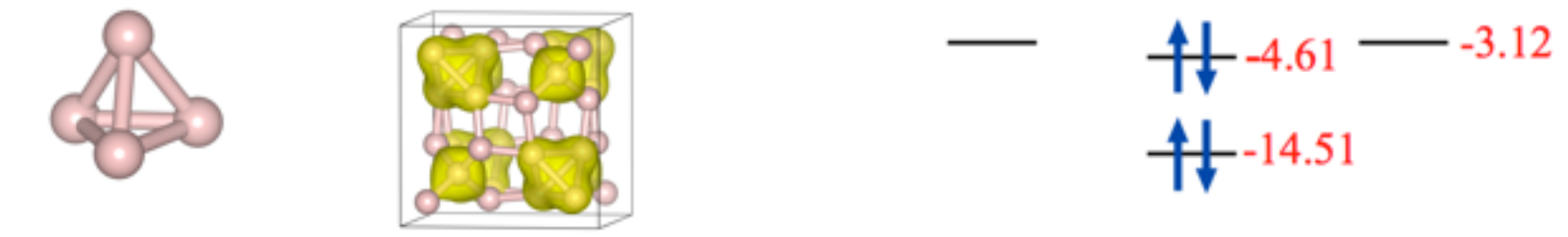  | | |
| $H_6$ hexagon | $MH_6$ M=Ca[4], Sc[5–7], Y[8,9] | 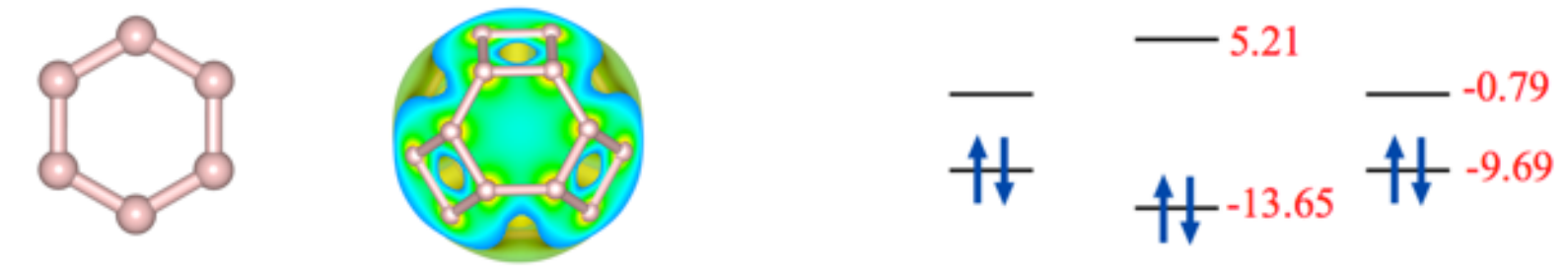  | | |
| $H_6$ corona | $MH_9$ M = Sc[1], Y[1], Th[10], Ce[11,12], Pr[2], U[13] | 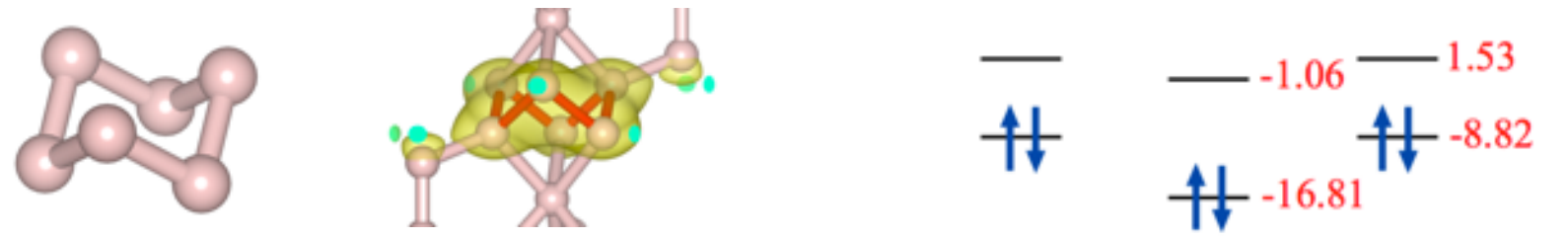  | | |
| $H_8$ cube | $MH_{10}$ M=La[1,14–17], Y[1], Th[10,18] | 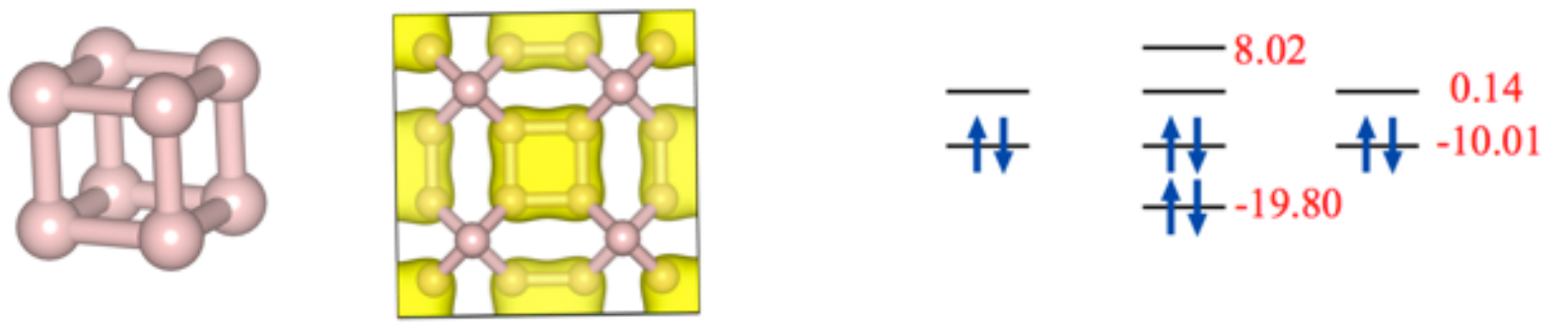  | | |
| $H_8$ deformed cube | $MH_9$ M=Pr[1,2], Pa[3] | 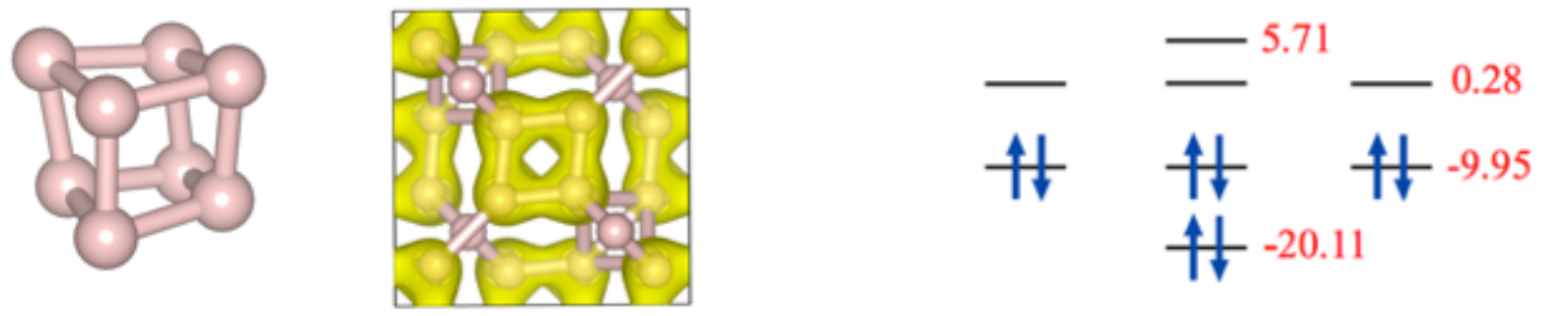  | | |
| $H_{10}$ Tripentagon | $MH_{10}$ M= Sc[19], Zr[19], Hf[19], Lu[19] | 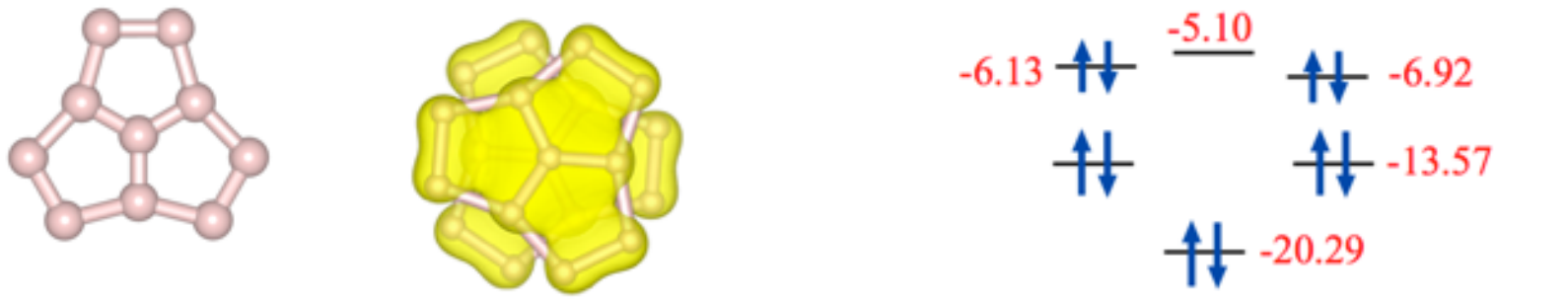  | | |
| $H_{12}$ Cuboctahedron | $MH_{12}$ M= Ba[20] | 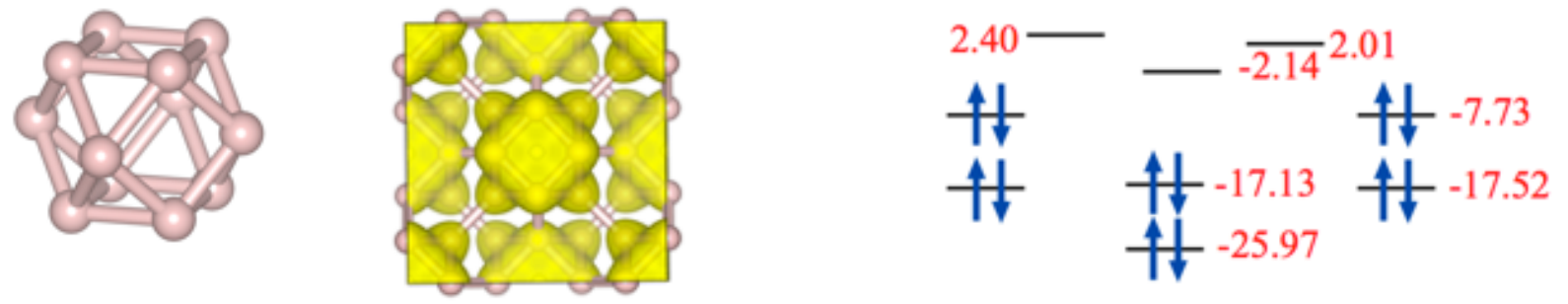  | | |
| $H_{12}$ Ring | $MH_{13}$ M=Y[1] | 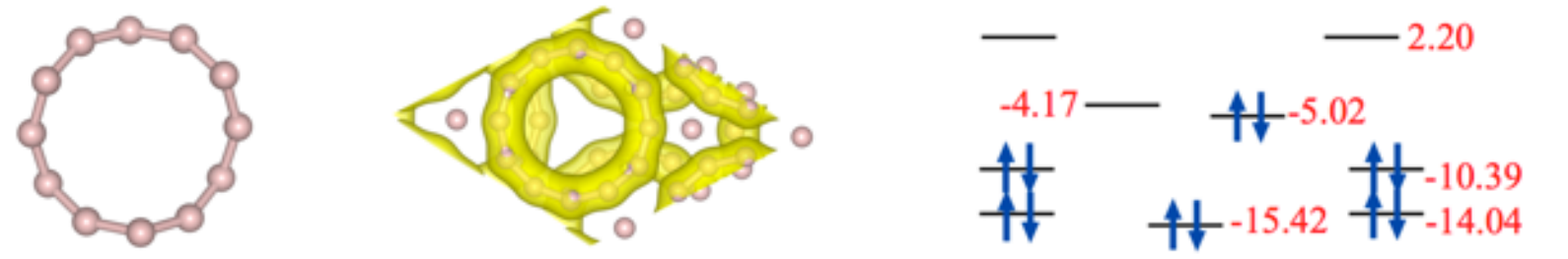  | | |
| $H_{14}$ Cage | $MH_{16}$ M = La[21] | 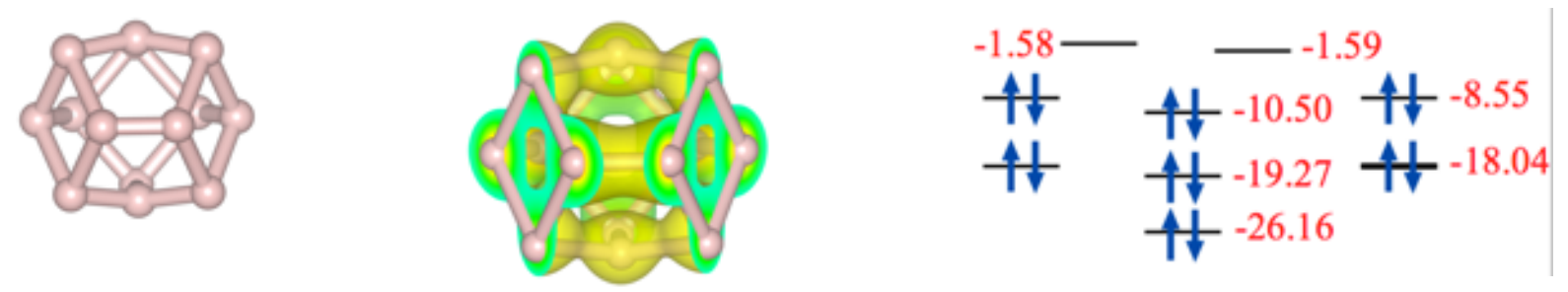  | | |

**Table 1.** The structures and the energy levels of H clusters (building units) in superhydrides. They are identified by geometries and wavefunctions of the H lattice after removing all metal atoms in superhydrides.

Another important structure feature of the superhydrides is that their H lattices consist of building units that show unique electronic structure features, including being aromatic. Aromaticity is a concept for organic molecules consisting of conjugated π bonds formed by the p orbitals of consisting atoms. The corresponding molecular orbitals are delocalized because each p orbital can bond resonantly with two or more neighboring p orbitals. The most common examples are those formed by C 2p orbitals. Because of the topological nature of the 2p orbitals, the C atoms need to locate in the same plane in aromatic molecules or groups. The essence of the aromaticity is that the energy levels of the π orbitals form a gap, below which all the levels are occupied and above which all the level are empty. Because of the coplanar constraint, the π orbitals come in pairs except the lowest energy one that corresponds to the overall bonding state and does not contain any node. Therefore, the π systems containing 4n+2 electrons are aromatic because they fill the 2n+1 orbitals and form a gap. The size of the gap corresponds to the strength of the aromaticity, and the larger the gap, the stronger the aromacity. In contrast to π bonds, the σ bonds are usually formed by $sp^n$ (n=3,2,1) hybrid orbitals. These bonds are localized in the region between the two bonding atoms because, each hybrid orbital can only form bond with another orbital in one direction (head to head). On the other hand, the $sp^3$ hybridized orbitals can form three-dimensional covalent networks, which becomes the structural framework for organic molecules and many covalent crystals.

The H-H bonds in superhydrides as well as in atomic hydrogen share the features of both the π bonds of p orbitals and σ bonds of hybrid orbitals. These bonds are formed by the 1s orbitals of two neighboring H atoms. Similar to σ bonds, they can form 3D networks as in superhydrides or in atomic hydrogen phases. On the other hand, one 1s orbital can form multiple bonds with neighboring H atoms resonantly in a similar way as the π bonds in organic molecules. However, because the 1s orbital is spherical, the conjugation of the 1s-1s bonds does not require these bonds to be coplanar. This major difference between the H-H bonds and the C-C π bonds lifts the coplanar constraint of aromaticity. Thus, a 3D H cluster could still be aromatic, if the occupation of its energy levels satisfies the requirement of aromaticity, i.e. the lower energy orbitals are fully occupied and the higher energy orbitals are empty, forming a gap between the HOMO and the LUMO orbitals. Eventually, the aromaticity depends on the symmetry and the number of electrons of the cluster. Strikingly, we find that the H lattices of most of the superhydrides consist of building units that are aromatic. Especially, these units locate at the high value regions of ELF, often corresponding to maxima, and therefore are largely stabilized by the electrons localized in these regions. The following summarizes the most important building units found in the superhydrides that have been discovered so far.

**$H_4$ tetrahedrons** are found in $F\bar{4}3m$ $MH_9$ (M=Pr etc.) compounds. The four electrons fill the two lower energy orbitals and the two degenerate levels with higher energy are empty. It has a gap of 1.49 eV if the geometry parameters are taken from $PrH_9$ at 400 GPa. In $MH_9$ superhydrides, the $H_4$ tetrahedrons locate at the tetrahedral interstitial sites where the ELF shows maxima.

**$H_6$ hexagons** are found in $Im\bar{3}m$ $MH_6$ superhydrides. Many metals such as Ca, Sc, Y and La etc. can form these superhydrides. The deformed $H_6$ hexagons, such as elongated $H_6$, can be found in several other types of superhydrides, such as $Fm\bar{3}m$ $MH_{10}$ and Immm $M_2H_{13}$ etc. The $H_6$ hexagon is analogous to benzene molecules. Its aromaticity has been thoroughly discussed previously and has been used to explain the stability of atomic hydrogen phases that consist of "graphene-like" hydrogen layered structures. The H-H distance in $H_6$ hexagons in $LaH_6$ at 300 GPa is 1.234 Å, and the corresponding HOMO-LUMO gap is found to be 8.50 eV, indicating a strong aromaticity.

**$H_6$ corona** is found in $P6_3/mmc$ $MH_9$ superhydrides. All 6 H atoms are equivalent. The H-H distance is 1.118 Å for $CeH_9$ at 300 GPa. It is also aromatic and has a large gap of 8.76 eV, similar to $H_6$ hexagon. Interestingly, the LUMO of $H_6$ corona is a single orbital, which is different to the doubly degenerate LUMO in $H_6$ hexagons. In $MH_9$ compounds, $H_6$ coronas locate at the octahedral interstitial sites and are significantly stabilized by the high distribution of electrons.

**$H_8$ cube** is found in $Fm\bar{3}m$ $MH_{10}$ superhydrides. The well-known example is $LaH_{10}$. The H-H distances in $H_8$ cube in $LaH_{10}$ at 300 GPa are 1.146 Å. This is actually longer than the $H^1$-$H^2$ distances that connect the $H_8$ cubes and the $H^2$ atoms at the tetrahedral sites. The later is 1.064 Å in $LaH_{10}$ at 300 GPa. Still, the wavefunctions show that $H_8$ cube is a building unit. $H_8$ cube show very strong aromaticity with a gap between its HOMO and LUMO states as high as 10.15 eV. Due to the highly symmetric geometry, both HOMO and LUMO orbitals are triply degenerate. In $MH_{10}$ superhydrides, $H_8$ cubes locate at the octahedral sites and are largely stabilized by the large electron localizations as revealed by the metal ELF.

**Deformed $H_8$ cube** presents in $F\bar{4}3m$ $MH_9$ superhydrides. It can be obtained by deforming the cube in a way that all faces become slightly bended rhombus. All H atoms are equivalent and there is only one nearest H-H distance. It has a value of 1.149 Å in $PrH_9$ at 400 GPa. The two angles of the rhombus are 78.04° and 100.80°, and the dihedral angle of the bended rhombus is 11.61°. With these geometry parameters, the energy levels of the deformed $H_8$ cube are calculated. Despite the lowering of the symmetry, the energy levels are similar to the original $H_8$ cube. Especially, the HOMO and LUMO orbitals are triply degenerate. It also shows a large gap of 10.23 eV, indicating strong aromaticity. Similar to $H_8$ cube, the deformed $H_8$ cubes also locate at the octahedral sites of FCC metal lattice and are greatly stabilized by the high electron distributions in these regions.

**$H_{10}$ tripentagons** are found in $P6_3/mmc$ $MH_{10}$ superhydrides. DFT calculations predicted the formation of such compounds for several metals including Sc, Zr, Lu and Hf, etc. While concerning the H lattices, this type of superhydrides is quite unique. First, the H atoms do not form an extended covalent network. Instead, they form planar $H_{10}$ molecules. The structure of these molecules is analogous to acepentalene ($C_{10}H_6$) and consists of three pentagons connected by sharing edges. There are three H-H distances. For $ScH_{10}$ at 300 GPa, the H-H distance between the center H and its three neighbors (H that are shared by two neighboring pentagons) is 0.915 Å, the H-H distance between the shared H and its neighboring edge H is 1.067 Å, and the H-H diatnce between two edge H atoms is 1.007 Å. The Gaussian calculation of the electronic structure of $H_{10}$ molecule with the same geometry as $H_{10}$ in $ScH_{10}$ at 300 GPa shows a small gap of 1.03 eV between its non-degenerate HOMO and LUMO levels.

**$H_{12}$ cuboctahedrons** exist in $Fm\bar{3}m$ $MH_{12}$ superhydrides. Comparing with cubic $MH_{10}$ structure, all H atoms locate at octahedral interstitials and form cuboctahedrons. In $BaH_{12}$ at 135 GPa, the shortest H-H distance within $H_{12}$ is 1.271 Å, whereas the shortest H-H distance between the neighboring $H_{12}$ is 1.344 Å. All H atoms are equivalent in cuboctahedrons. With the above geometry, the energy levels are calculated using Gaussian program. The HOMO levels are doubly degenerate whereas the LUMO is not. The gap between the HOMO and the LUMO levels is found to be 5.59 eV, showing moderate aromaticity. Like all other H clusters, the $H_{12}$ cuboctahedrons are also stabilized by the electron distributions at the octahedral sites in metal FCC lattices.

**$H_{12}$ ring** is a striking unit found in $R\bar{3}m$ $MH_{13}$ compounds. Its large size (3.900 Å diameter in $YH_{13}$ at 200 GPa) and unusual shape make it difficult for close packing under high pressure. However, both the geometry and the wavefunctions distinctly reveal their existence. For $YH_{13}$ at 200 GPa, the H-H bond length within the ring is 1.025 Å, whereas the shortest H-H distance between the rings is 1.224 Å and the shortest distance between H on the ring and the H between the layers is 1.210 Å. Different to other H units, $H_{12}$ ring is an analogue to C π bonds, therefore should also follow the Hückel 4n+2 rule. Based on this rule, $H_{12}$ ring is anti-aromatic, i.e. the doubly degenerate HOMO levels are half filled. However, in the $YH_{13}$ structure, the $H_{12}$ ring is slightly distorted. Among 12 H atoms, 6 remain in the original position, 3 move upwards and 3 downwards alternatively, creating a wave-like modulation. As the result of this deformation, a gap of 0.85 eV is opened. As discussed in the previous section, $H_{12}$ ring matches the ELF high value regions of the metal lattice and therefore its stability is enhanced by the template effect.

**$H_{14}$ cage** can be viewed as the structure unit of the H lattice in P6/mmm $MH_{16}$ superhydrides. Not all the H atoms are equivalent and there are three different H-H distances in the cage. The H-H distance of the top H to its three neighbors (second layer) is 1.074 Å, and the H-H distance that connects the second and the third layer H atoms is 1.271 Å. Lastly, the two neighboring H atoms in the third layer has a distance of 1.011 Å. The unit is aromatic and shows a gap of 6.96 eV between its doubly degenerate HOMO levels and the single LUMO level. In $MH_{16}$ structures, the $H_{14}$ cages are connected by sharing the $H_4$ rings at its three sides, and form a 2D layered structure with hexagonal symmetry. The 2D hexagonal lattice matches well with the ELF of the metal lattice and are stabilized by the electrons localized in the region.

# Section III. The chemical templates in various superhydrides

| Compound (symmetry) | Metal Lattice | Metal ELF overlay on H lattice |
|---|---|---|
| $MH_6$ ($Im\bar{3}m$) | BCC<br>M=Ca[4], Sc[5–7], Y[8,9] | 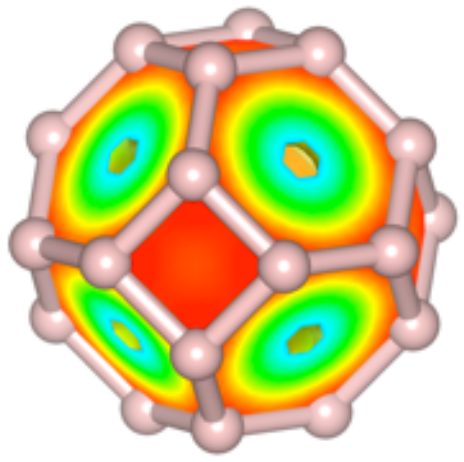 |
| $MH_{10}$ ($Fm\bar{3}m$) | FCC<br>M=La[1,14–17], Y[1], Th[10,18] | 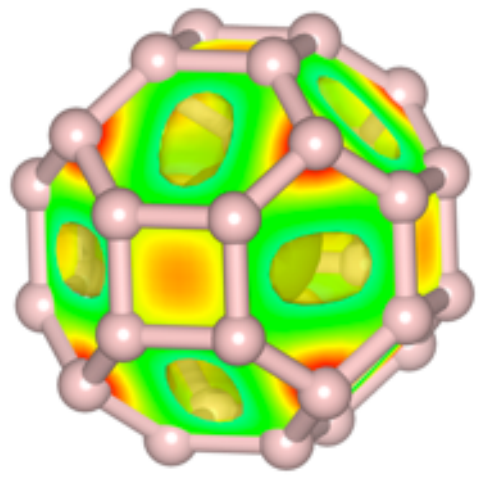 |
| $MH_9$ ($P6_3/mmc$) | HCP<br>M = Sc[1], Y[1], Th[10], Ce[11,12], Pr[2], U[13] | 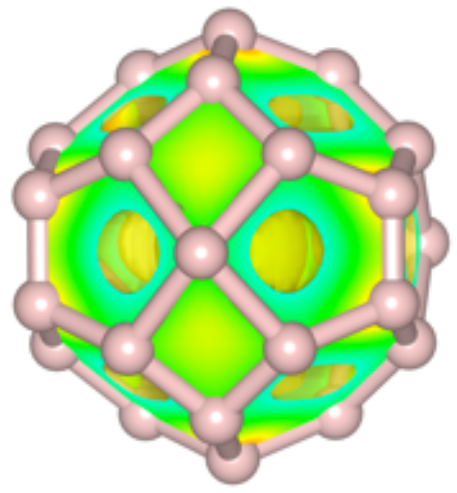 |
| $MH_6$ ($R\bar{3}m$) | dBCC*<br>M = Sr[22,23] | 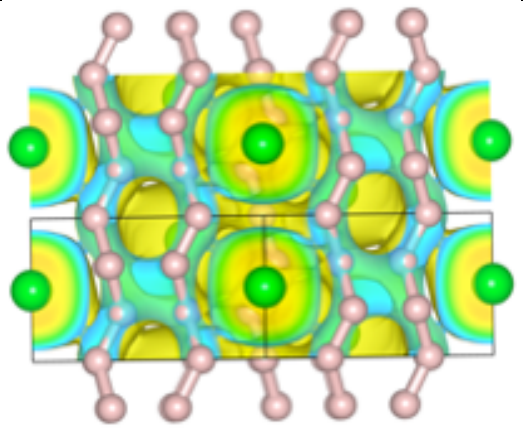 |
| $M_2H_{13}$ ($Immm$) | M=Ti[24] | 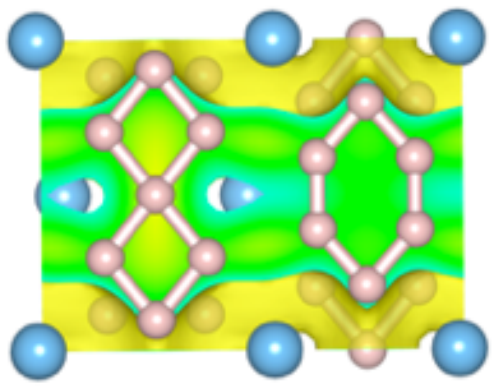 |
| $MH_9$ ($F\bar{4}3m$) | FCC<br>M=Pr[1,2], Pa[3] | 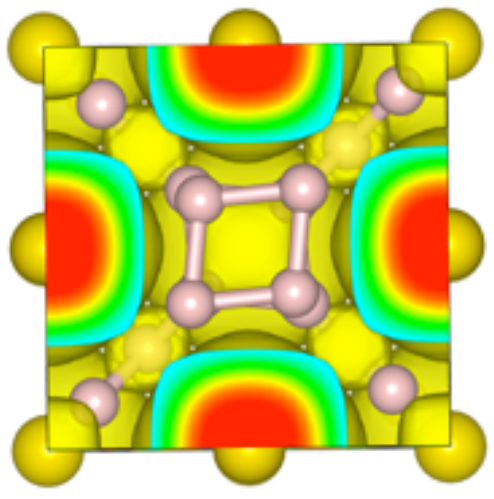 |
| $MH_{10}$ ($P6_3/mmc$) | HCP<br>M= Sc[19], Zr[19], Hf[19], Lu[19] | 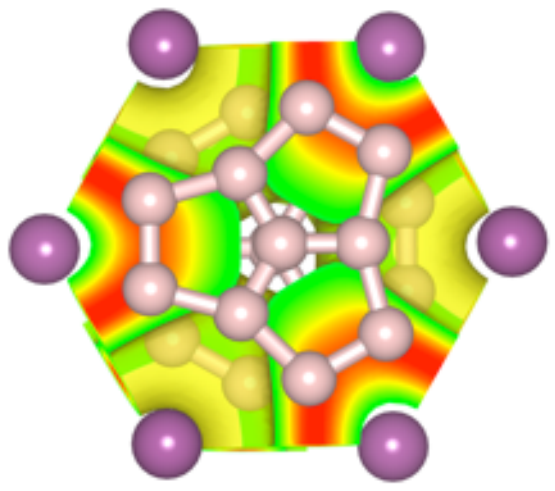 |
| $MH_{10}$ ($R\bar{3}m$) | Hex**<br>M = Ac[25] | 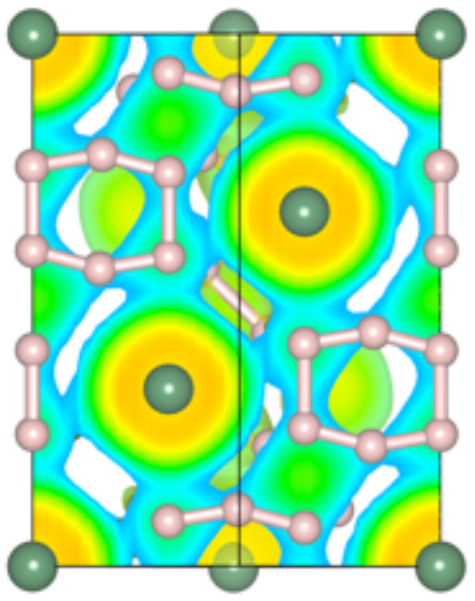 |
| $MH_{11}$ (P4nmm) | BCT<br>M= La[1,21] | 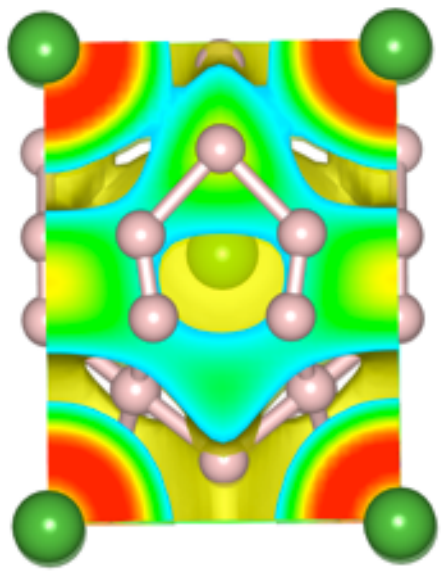 |
| $MH_{12}$ ($Fm\bar{3}m$) | FCC<br>M= Ba[20] | 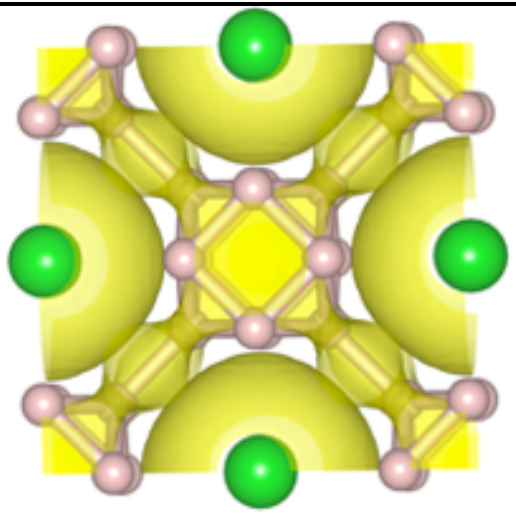 |
| $MH_{13}$ ($R\bar{3}m$) | FCC R-3m<br>M=Y[1] | 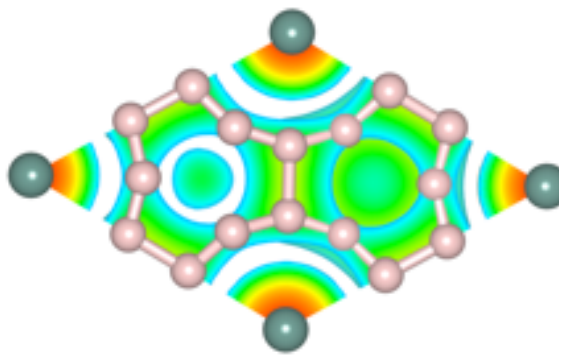 |
| $MH_{16}$ ($P6/mmm$) | SH<br>M = La[21] | 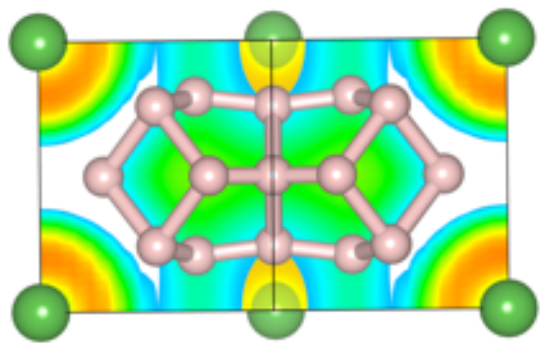 |

**Table 2.** The ELFs are calculated for metal lattices in superhydrides. The results are overlaid on H lattice. The small white balls represent H atoms, and the metal atoms, if visible, are represented by colored balls. *dBCC is distorted BCC. **Hex is an FCC deformed along (111).

**$Fm\overline{3}m$ LaH$_{10}$.** The chemical template mechanism is based on a striking feature of superhydrides that are emerged from electronic structure calculations. It can be revealed by calculating the ELF of the metal lattices in superhydrides after removing all H atoms. For example, the La atoms form an FCC lattice in $LaH_{10}$ superhydrides. For $LaH_{10}$ at 300 GPa, the La lattice corresponds to an FCC La at 12.4 GPa. The ELF of this metal structure is calculated and found to have large values and maxima at octahedral and tetrahedral interstitial sites. ELF is known to reveal the chemical structure of atoms, molecules and solids. For example, it can clearly reveal the shell structure and the orbitals of heavy atoms whereas the electron density decreases monotonically with increasing radial distance. Thus, the ELF maxima at the interstitial sites of FCC La indicate that some crystal orbitals have large contributions in these regions. In other words, these crystal orbitals consist of large compositions of the local orbitals of the quasi-atoms locating at these interstitial sites. As discussed in the text, we can identify the crystal orbitals that correspond to the occupation of quasi-atom orbitals at the interstitial sites. Since the electrons do not localize completely at these interstitial sites, it is not a high-pressure electride (HPE).

The large electron distributions at the interstitial sites act as an effective chemical template for the assembly of H lattice in $LaH_{10}$ because they match nicely with the $H_{10}$ lattice. As shown in Fig. 2a and 4b, the $H_8$ cubes and the $H^2$ atoms of the H lattice in $LaH_{10}$ locate exactly at the $E^O$ and $E^T$ interstitial sites.

**$Im\overline{3}m$ MH$_6$ (M=Ca, Sc, Y etc.)** The metal atoms form a BCC lattice. The ELF shows large values at the octahedral interstitial sites. Once overlaid on the $H_6$ lattice, it shows large values on the sites of the H atoms, especially it matches well with the $H_4$ and $H_6$ rings in the H lattice.

**$P6_3/mmc$ MH$_9$ (M=Ce, Y etc.)** The metal atoms form an HCP lattice. The ELF has maxima at the octahedral and tetrahedral interstitial sites. Again, it matches well with the H lattice in $MH_9$ superhydrides as shown in Fig. 4c and Supplementary Table 2. The $H_6$ corona locates at the octahedral sites whereas the two center H atoms in $H_8$ bipyramid units locate close to the neighboring tetrahedral sites.

**$R\overline{3}m$ MH$_6$ (M=Sr)** The metal atoms form a deformed BCC lattice ($R\overline{3}m$). It can be viewed as compressing an BCC lattice along the (111) diagonal direction. The ELF shows maxima at metal sites as well as the interstitial sites. With an isosurface value of 0.3, the ELF surface forms a 3D structure consisting of connected wavy chains along the perpendicular direction. Strikingly, this connected chain structure matches very well the H lattice in $SrH_6$, which consists of zigzag H chains. The H-H distances in the chain is 1.015 Å at 250 GPa, whereas the shortest H-H distances between the neighboring chains is 1.411 Å.

***Immm*** **$M_2H_{13}$ (M=Ti)** This superhydride is found to adopt a layered structure. In each layer, the H atoms form two ribbons. One ribbon consists of a chain of double-rhombus; and the other consists of isolated elongated hexagons. The H-H distances in these elongated hexagons are 1.023 and 1.048 Å. The H-H distances between two hexagons are 1.815 Å, ruling out the connection by a covalent bond. In contrast, the H-H distance between the double-rhombus is only 0.865 Å, indicating the presence of $H_2$ molecules. This ribbon can also be viewed as connected $H_2$ molecules and square planar $H_5$ units. Despite the complicated structure of H lattice, it matches very well the ELF maxima of the metal lattice. As shown in Supplementary Table 2, the ELF also show layered feature. Especially, it shows large values in the regions where the double-rhombus and the elongated $H_6$ hexagons locate. Interestingly, the locations of the $H_2$ molecules correspond to regions of minimum ELF.

**$F\bar{4}3m$ $MH_9$ (M=Pr)** The structures can be viewed as a deformed $LaH_{10}$ structure. Although the symmetry is lower than that of $MH_{10}$, the metal lattice maintains an FCC structure. Therefore, its ELF is the same as in $MH_{10}$, and show maxima at both the octahedral and the tetrahedral interstitial sites. There are two major differences between the $H_9$ lattice in this compound and the $H_{10}$ lattice in $MH_{10}$. First, half of the $H^2$ atoms are replaced by $H_4$ tetrahedrons at the tetrahedral sites. Second, the $H_8$ cubes are deformed to match the lower symmetry caused by the replacement of $H^2$ atoms by $H_4$ tetrahedrons. Nevertheless, the $H_9$ lattice matches well with the ELF, with deformed $H_8$ cubes locating at the octahedral sites and the $H^2$ atoms and $H_4$ tetrahedrons occupying the tetrahedral sites.

**$P6_3/mmc$ $MH_{10}$ (M=Sc, Zr, Lu, Hf etc.)** Different to other $MH_{10}$ superhydrides, these compounds adopt a hexagonal layered structure ($P6_3/mmc$). Each layer contains 1 metal atom and 10 H atoms positions in a hexagonal lattice. Another important feature is that the H in these compounds do not form extended covalent network. Instead, they form $H_{10}$ planar molecules with the same geometry of acepentalene ($C_{10}H_6$). The metal lattice is an HCP reduced in perpendicular direction. The ELF of this metal lattice shows rings of large value maxima locating at the B sites (if the metals locate at A sites). The highest values at the interstitial regions are as large as 0.65 for Sc. The $H_{10}$ molecule locate in between the ELF rings, with three of its edges (2 H atoms) locating in the high ELF value region, and the H atoms sharing two pentagons locating in the medium ELF value regions. The center H atoms locate at ELF minima.

**$R\bar{3}m$ $MH_{10}$ (M=Ac etc.)** The metal lattice of this superhydride is very close to an FCC lattice. It is slightly compressed along the (111) diagonal direction. The M-M distances inside the [111] layer are 3.688 Å whereas the M-M distances between neighboring layers are 3.497 Å. Nevertheless, the H lattice is very different to cubic $MH_{10}$. It is a semi-layered structure. The in-layer H-H distances are 1.085 and 1.099 Å, whereas the shortest H-H distances between layers are 1.313 Å. Despite the complex structure of H lattice, it matches well with the ELF. Similar to the ELF of an FCC structure, the ELF results of the metals in these compounds also show maxima in the octahedral and the tetrahedral interstitial sites. A deformed $H_8$ cube locates at each octahedral site, and one H atom occupies each tetrahedral site.

***P4nmm* $MH_{11}$ (M=La etc.)** The metal atoms form a body-centered tetragonal (BCT) lattice. The lattice parameters at 100 GPa are a=b=3.873 Å and c=5.272 Å. The major interstitial sites are octahedral and locate at the base center. ELF of BCT La shows maxima at these sites. The H lattice matches fairly well with the ELF distribution.

**$Fm\bar{3}m$ $MH_{12}$ (M=Ba etc.)** This type of superhydrides has the same symmetry group ($Fm\bar{3}m$) as cubic $LaH_{10}$. Therefore, its metal atoms form an FCC lattice. The ELF shows maxima at both the octahedral and interstitial sites. The major difference is the H lattice. The H lattice in $MH_{12}$ consists of $H_{12}$ cuboctahedrons (polyhedrons with 12 vertices). They all locate at the octahedral sites of the metal FCC lattice. There is no H atom locating at the tetrahedral sites.

**$R\bar{3}m$ $MH_{13}$ (M=Y etc.)** Similar to all other $R\bar{3}m$ superhydrides ($MH_6$ and $MH_{10}$), the metal atoms form a deformed FCC by compressing the lattice along (111) diagonal direction. The H lattice shows a very unusual structural feature. Among the 13 H atoms, 12 of them locate in the same layer with the metal atom, forming a large ring circling the metal atom. The H-H distances in the ring are 1.025 Å for $YH_{13}$ under 200 GPa, the shortest H-H distance between rings is 1.224 Å. One extra H locates between the two layers and at the center site among three $H_{12}$ rings, and connects them with a H-H distance of 1.210 Å. The high value regions of ELF match the H lattice including the $H_{12}$ rings in the layers and the H atoms connecting the rings.

***P6/mmm* $MH_{16}$ (M=La)** The metal atoms in this superhydrides form a simple hexagonal (SH) lattice. The lattice parameters for both $MH_{16}$ at 150 GPa and the corresponding SH lattice are a=b=3.692 Å and c=3.707 Å. The H lattice consists of layers. Each layer is formed by connecting $H_{14}$ cages by their 3 $H_4$ rhombus sides. There are three H-H distances inside each $H_{14}$ cage that are 1.074, 1.011 and 1.271 Å. The minimum H-H distance between two neighboring layers is 1.525 Å. The ELF of the metal lattice show maxima at the two interstitial sites centered at the B and C sites of hexagonal lattice (assuming the metal atoms locate at the A sites) and at the center of the two metal layers. It matches very nicely with the H lattice. Each $H_{14}$ cage centers at an interstitial site.

# Section IV. The normal stresses of the metal sub-lattices in superhydrides

| | Compounds | Pressure (GPa) | Metal Lattice Pressure (GPa) | | | | Compounds | Pressure (GPa) | Metal Lattice Pressure (GPa) | | |
|---|---|---|---|---|---|---|---|---|---|---|---|
| $MH_6$ ($Im\bar{3}m$) | $CaH_6$ | 300 | 56.2 | 56.2 | 56.2 | $MH_{10}$ ($P6_3/mmc$) | $ScH_{10}$ | 300 | 9.3 | 9.3 | 30.7 |
| | $ScH_6$ | 300 | 37.0 | 37.0 | 37.0 | | $ZrH_{10}$ | 300 | 6.5 | 6.5 | 26.7 |
| | $YH_6$ | 300 | 45.8 | 45.8 | 45.8 | | $LuH_{10}$ | 300 | 13.8 | 13.8 | 49.4 |
| | | | | | | | $HfH_{10}$ | 300 | 5.2 | 5.2 | 33.7 |
| $MH_{10}$ ($Fm\bar{3}m$) | $LaH_{10}$ | 300 | 12.4 | 12.4 | 12.4 | $MH_{10}$ ($R\bar{3}m$) | $LaH_{10}$ | 150 | 2.1 | 2.1 | 4.7 |
| | $YH_{10}$ | 300 | 16.4 | 16.4 | 16.4 | | $AcH_{10}$ | 200 | 11.3 | 11.3 | 18.7 |
| | $ThH_{10}$ | 300 | 10.1 | 10.1 | 10.1 | | | | | | |
| $MH_9$ ($P6_3/mmc$) | $ScH_9$ | 300 | 11.0 | 11.0 | 12.0 | $MH_{11}$ (P4nmm) | $LaH_{11}$ | 100 | -1.1 | -1.1 | -0.6 |
| | $YH_9$ | 300 | 22.0 | 22.0 | 24.1 | | | | | | |
| | $ThH_9$ | 300 | 11.9 | 11.9 | 35.1 | | | | | | |
| | $CeH_9$ | 300 | 1.1 | 1.1 | 15.1 | | | | | | |
| | $PrH_9$ | 300 | -3.2 | -3.2 | 8.8 | | | | | | |
| | $UH_9$ | 300 | -17.4 | -17.4 | -5.0 | | | | | | |
| $MH_6$ ($R\bar{3}m$) | $SrH_6$ | 300 | 49.2 | 49.2 | 59.0 | $MH_{12}$ ($Fm\bar{3}m$) | $BaH_{12}$ | 135 | 6.3 | 6.3 | 6.3 |
| $M_2H_{13}$ ($Immm$) | $Ti_2H_{13}$ | 350 | 23.1 | -6.3 | 21.5 | $MH_{13}$ ($R\bar{3}m$) | $YH_{13}$ | 200 | 4.1 | 4.1 | -7.7 |
| $MH_9$ ($F\bar{4}3m$) | $PrH_9$ | 400 | 4.9 | 4.9 | 4.9 | $MH_{16}$ ($P6/mmm$) | $LaH_{16}$ | 150 | -3.5 | -3.5 | -3.4 |
| | $PaH_9$ | 300 | -1.2 | -1.2 | -1.2 | | | | | | |

**Table 3.** Normal stresses of the metal lattices (after removing all H atoms) in superhydrides under certain pressures.

## Section V. The chemical templates strengths in $MH_6$, $MH_{10}$ and $MH_9$

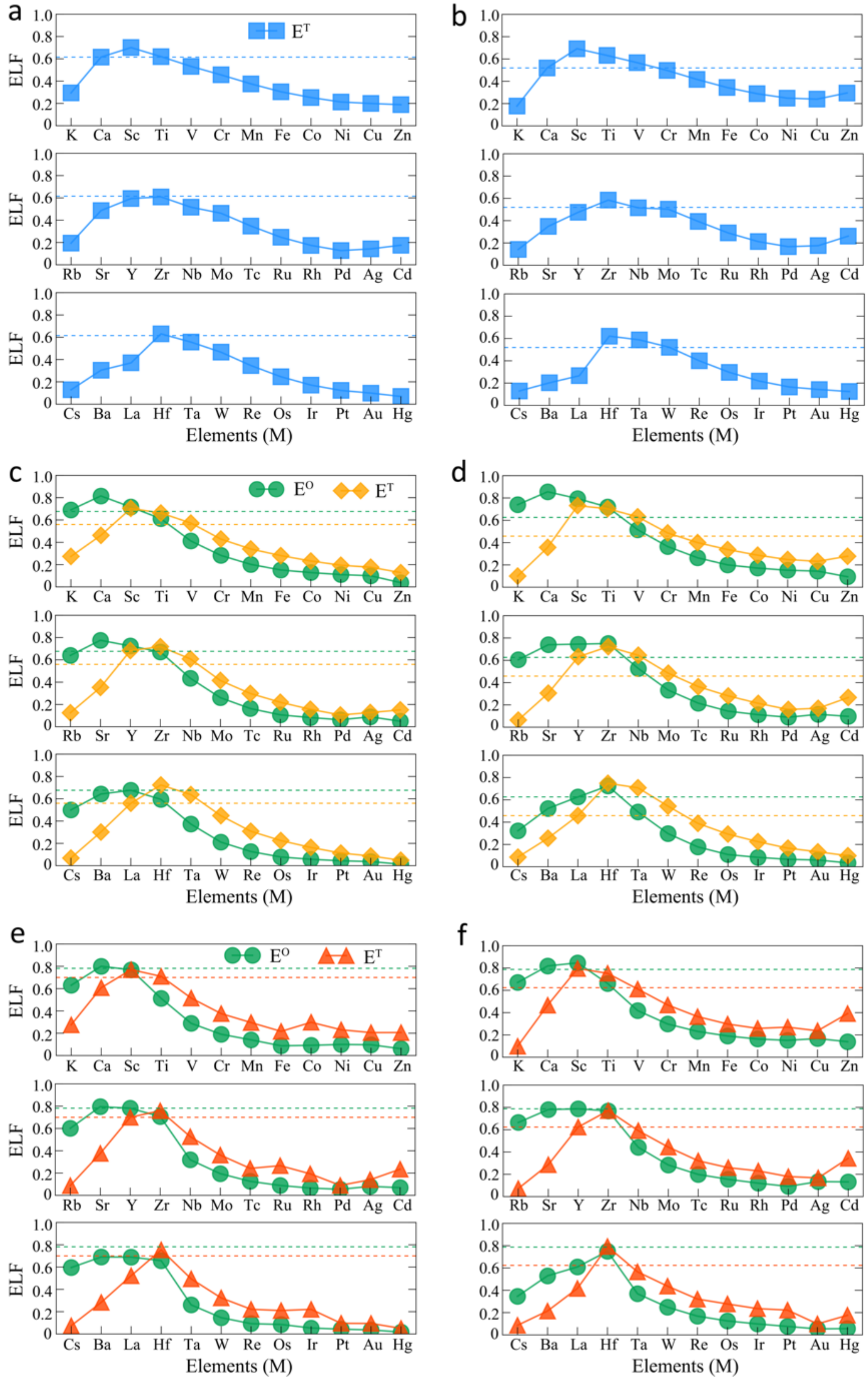

**Fig. 2: The ELF values at the interstitial points of metal lattices in $MH_6$, $MH_{10}$ and $MH_9$. a**, The ELF values at $E^T$ sites of metal BCC lattice with a unit length corresponds to $MH_6$ superhydrides at 100 GPa. M is a 4th, 5th, and 6th row metal element, respectively. **b**, The ELF values at $E^T$ sites of metal BCC lattice with a unit length corresponds to $MH_6$ superhydrides at 300 GPa. The blue dashed lines in both (**a**) and (**b**) panels show the ELF values of Ca for comparison. **c**, The ELF values at $E^O$ and $E^T$ sites of metal FCC lattice with a unit length corresponds to $MH_{10}$ at 100 GPa. M is a 4th, 5th, and 6th row metal element, respectively. **d**, The ELF values at $E^O$ and $E^T$ sites of metal FCC lattice with a unit length corresponds to $MH_{10}$ at 300 GPa. The green and orange dashed lines show the ELF values of La at $E^O$ and $E^T$ for comparison. **e**, The ELF values at $E^O$ and $E^T$ sites of metal HCP lattice with a unit length corresponds to $MH_9$ at 100 GPa. M is a 4th, 5th, and 6th row metal element, respectively. **f**, The ELF values at $E^O$ and $E^T$ sites of metal HCP lattice with a unit length corresponds to $MH_9$ at 100 GPa. The green and red dashed lines show the ELF values of Ce at $E^O$ and $E^T$ for comparison.

# Section VI. Stabilization of H units in superhydrides lattice

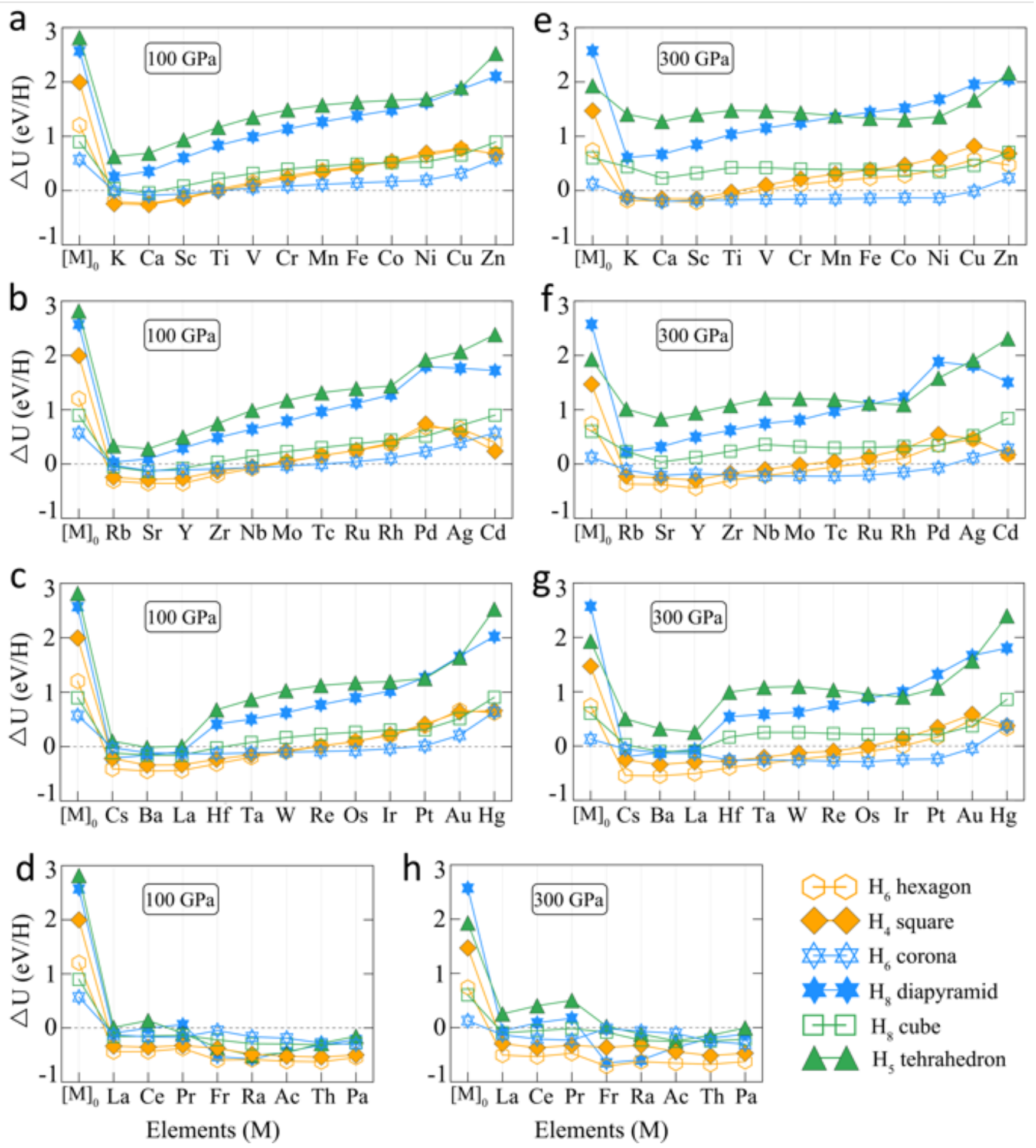

**Fig. 3: The change of internal energies while forming H clusters in metal superhydrides. a-d**, The energies of the H units relative to $H_2$ with the presence of metals in the 4$^{th}$ row (panel **a**), the 5$^{th}$ row (panel **b**) of periodic table, the 6$^{th}$ row (panel **c**) of periodic table and several rare earth metals (panel **d**), calculated by use of the He matrix at 100 GPa (see Methods). $[M]_0$ represents He matrix with no metals. **e-h**, The energies of the H units relative to $H_2$ with the presence of metals in the 4$^{th}$ row (panel **e**), the 5$^{th}$ row (panel **f**) of periodic table), the 6$^{th}$ row (panel **g**) of periodic table and several rare earth metals (panel **h**), calculated by use of the He matrix at 300 GPa.

## Section VII. Charge transfers in views of chemical bonds and chemical templates

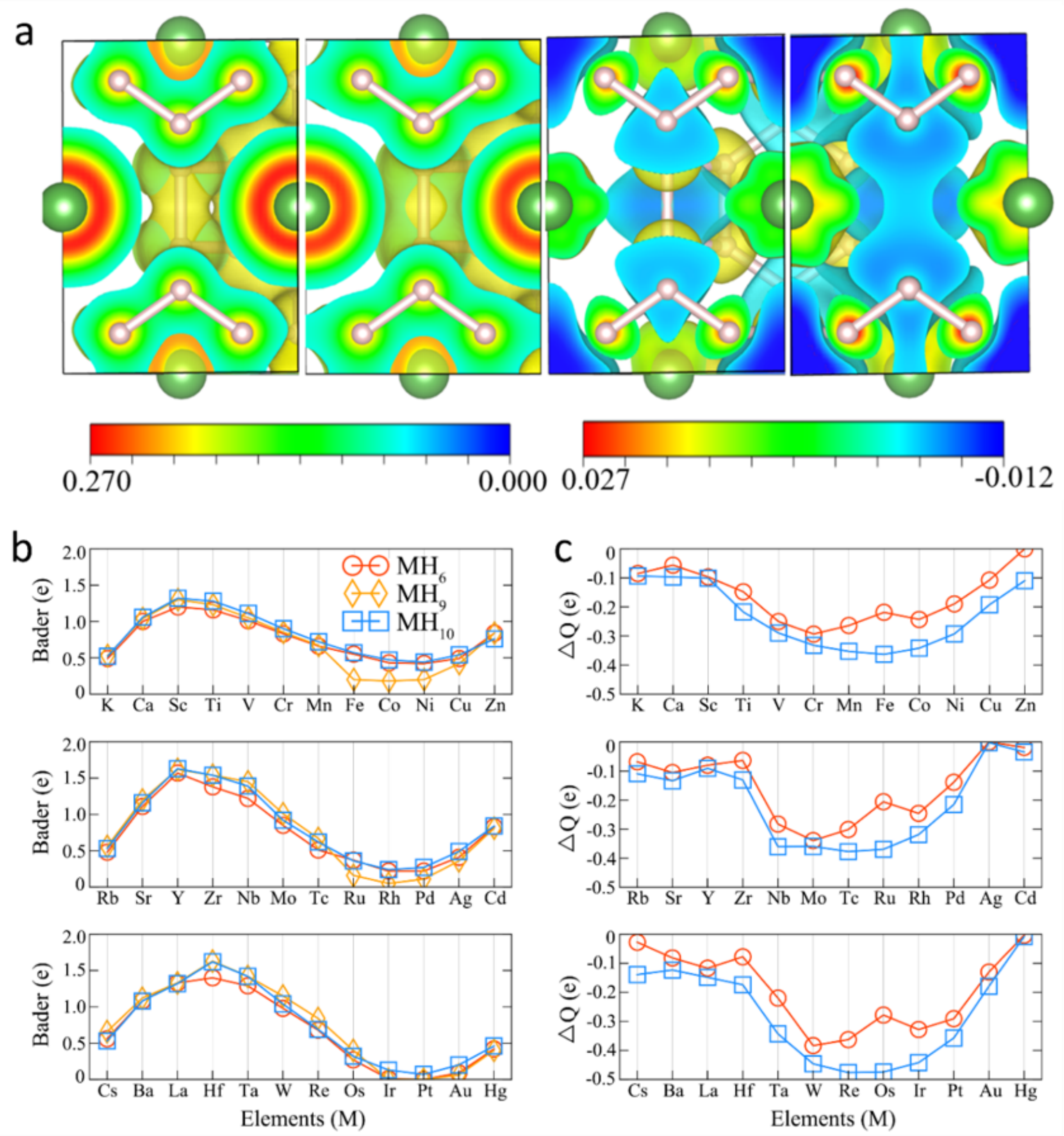

**Fig. 4: The charge transfers calculated using free atoms or sub-lattices as the reference states.** **a**, From left to right: the charge distribution of $LaH_{10}$ [$\rho(LaH_{10})$] on (110) plane at 100 GPa; the summation of the charge distributions of La [$\rho(La)$] and $H_{10}$ lattices [$\rho(H_{10})$] in $LaH_{10}$ compound on (110) plane at 100 GPa; the difference between $\rho(LaH_{10})$ and $\rho(La) + \rho(H_{10})$ at 100 GPa; the difference between $\rho(LaH_{10})$ and $\rho(La) + \rho(H_{10})$ at 300 GPa. **b**, The Bader charges of $MH_6$, $MH_{10}$ and $MH_9$ superhydrides for the 4$^{th}$, 5$^{th}$ and 6$^{th}$ rows of metal elements. **c**, The integrated differential charge density of metals in $MH_6$, $MH_{10}$ and $MH_9$ superhydrides (see Methods).

## Section VIII. The chemical templates and stability of mixed metal superhydrides

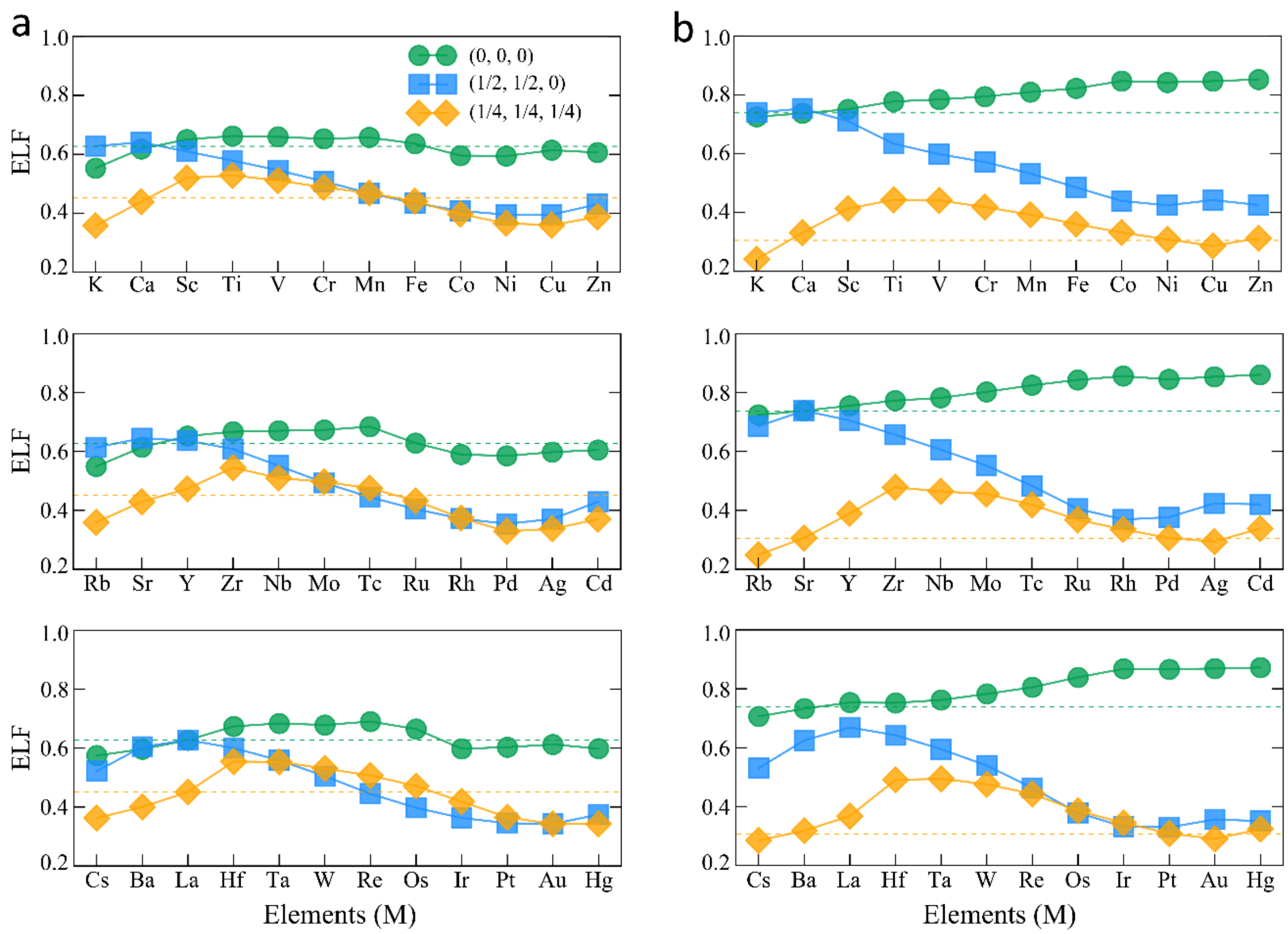

**Fig. 5: The change of ELF values while replacing metals in $MH_{10}$. a**, The change of ELF values at interstitial sites including (0, 0, 0), (1/2, 1/2, 0) and (1/4, 1/4, 1/4), while replacing one La atom in a conventional cell of $LaH_{10}$ superhydrides at 100 GPa. M is a $4^{th}$, $5^{th}$, and $6^{th}$ row metal element, respectively. The green and the orange dashed lines show the ELF values of La FCC at (0,0,0) and (1/4, 1/4, 1/4) points for comparison. **b**, The change of ELF values at interstitial sites including (0, 0, 0), (1/2, 1/2, 0) and (1/4, 1/4, 1/4), while replacing one Sr atom in a conventional cell of $SrH_{10}$ superhydrides at 100 GPa. M is a $4^{th}$, $5^{th}$, and $6^{th}$ row metal element, respectively. The green and the orange dashed lines show the ELF values of Sr FCC at (0,0,0) and (1/4, 1/4, 1/4) points for comparison.

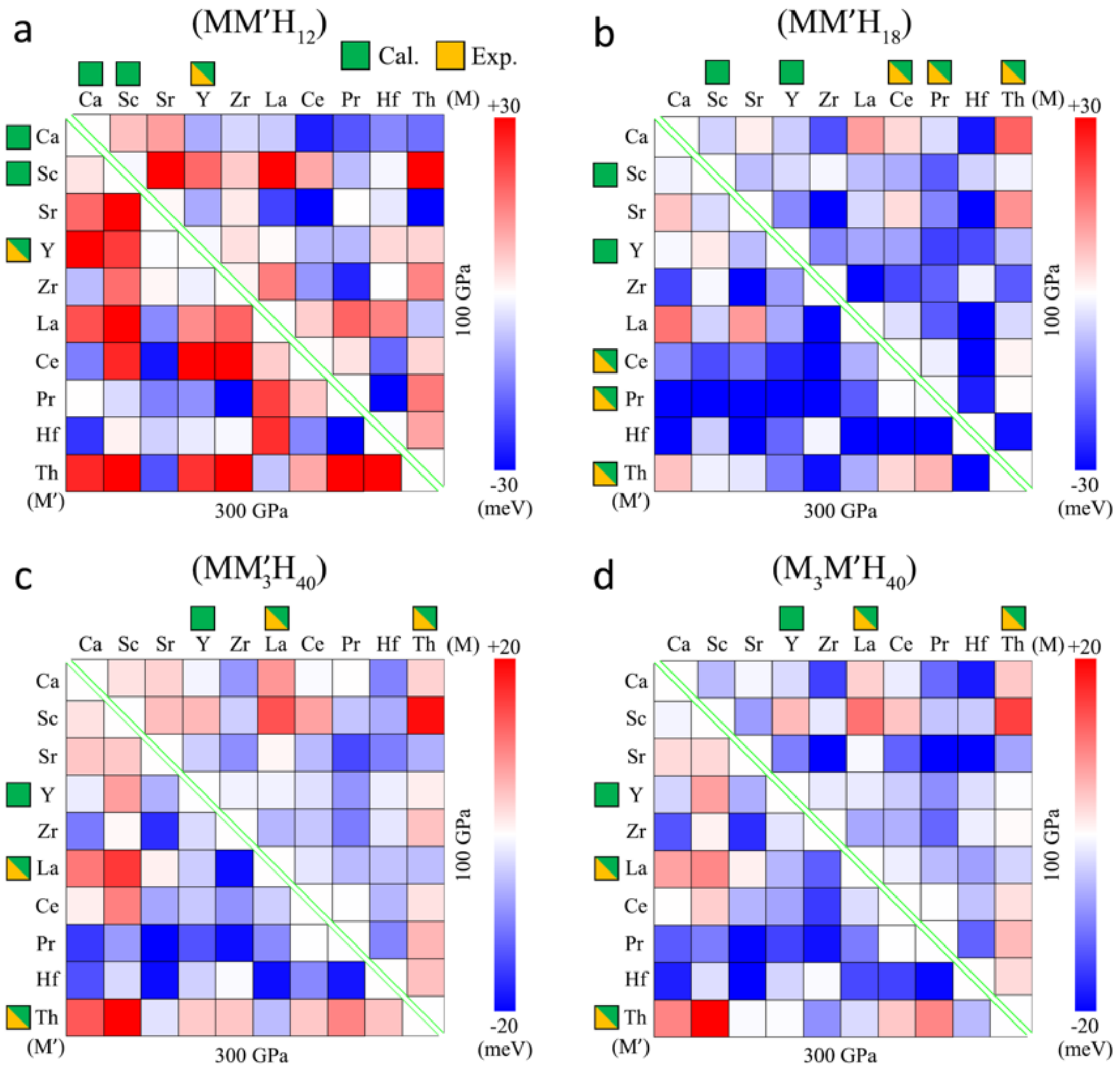

**Fig. 6: The reaction enthalpies of two metal superhydrides with the same H lattice forming a mixed metal superhydrides.** The formulas for calculating the reaction energies are shown in the Methods section of the text. The energies are shown in meV per atom. Negative reaction enthalpies mean the two metal superhydrides are more stable. **a**, The reaction energies of forming $MM'H_{12}$ from $MH_6$ and $M'H_6$. **b**, The reaction energies of forming $MM'H_{18}$ from $MH_9$ and $M'H_9$. **c**, The reaction energies of forming $MM'_3H_{40}$ from $MH_{10}$ and $M'H_{10}$. **d**, The reaction energies of forming $M_3M'H_{40}$ from $MH_{10}$ and $M'H_{10}$.

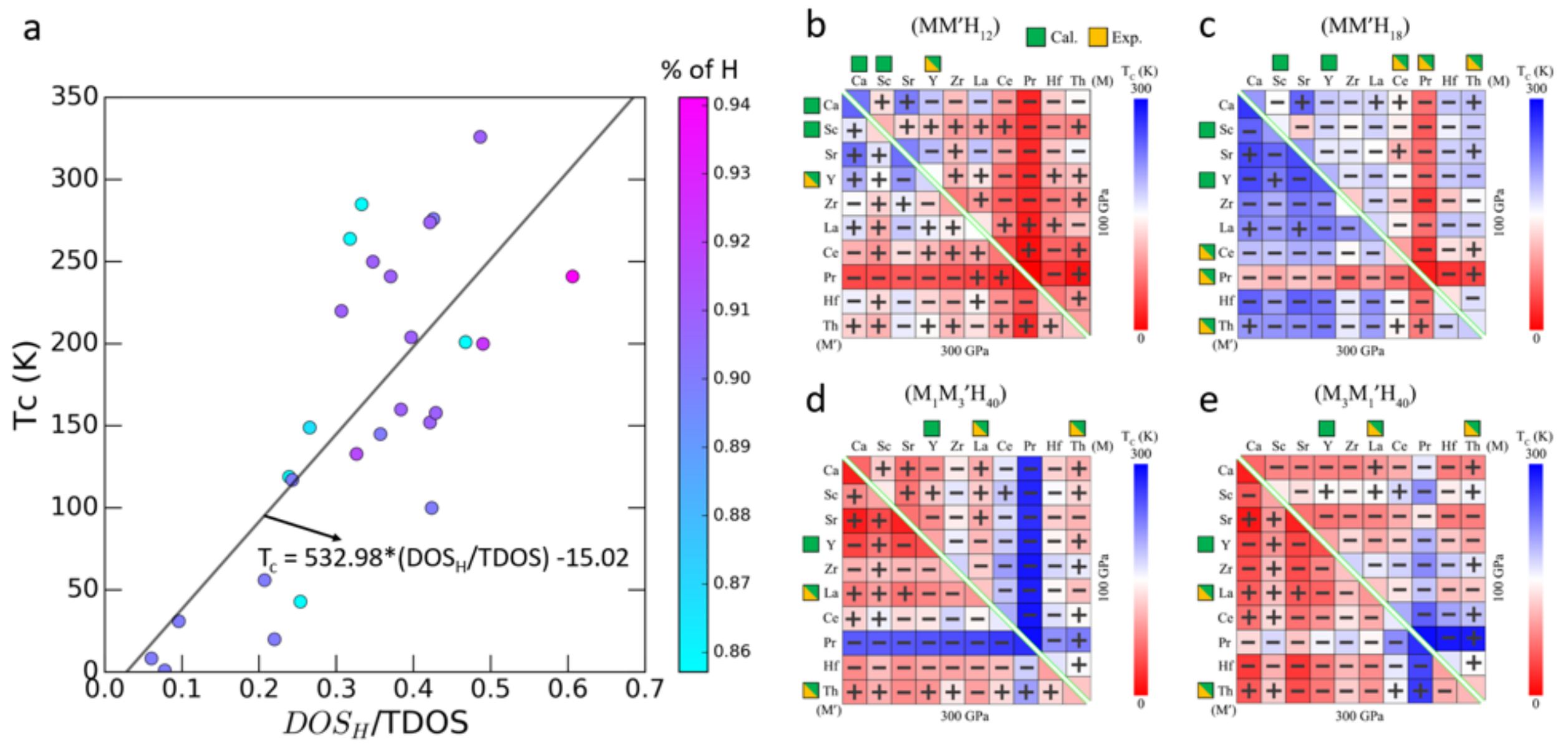

**Fig. 7: Estimate $T_c$ for mixed metal superhydrides. a**, The relation between $T_c$ and the ratio of density of hydrogen states ($DOS_H$) and total density of states (TDOS) at Fermi level of metal superhydrides predicted by DFT method. The $T_c$ values are taken from literatures. **b – e**, The estimated $T_c$ using calculated $DOS_H$/TDOS at Fermi level for the conceived mixed metal superhydrides. The order of the metals and rule of combinations are kept the same as in Supplementary Fig. 6. The color shows $T_c$ (blue represents higher $T_c$), whereas the + and - signs show the sign of the reaction enthalpies that are presented in Supplementary Fig. 6, *i.e.* - signs indicate that the mixed metal superhydrides are stable against the decomposition into single metal superhydrides. These results show that mixing metals in superhydrides has the potential to improve $T_c$ by increasing the density of hydrogen related states at the Fermi level.